# Review and Analysis of Networking Challenges in Cloud Computing

JOSE MOURA, Instituto Universitário de Lisboa (ISCTE-IUL), Instituto de Telecomunicações, Portugal
DAVID HUTCHISON, Lancaster University, Infolab21, UK

Cloud Computing offers virtualized computing, storage, and networking resources, over the Internet, to organizations and individual users in a completely dynamic way. These cloud resources are cheaper, easier to manage, and more elastic than sets of local, physical, ones. This encourages customers to outsource their applications and services to the cloud. The migration of both data and applications outside the administrative domain of customers into a shared environment imposes transversal, functional problems across distinct platforms and technologies. This article provides a contemporary discussion of the most relevant functional problems associated with the current evolution of Cloud Computing, mainly from the network perspective. The paper also gives a concise description of Cloud Computing concepts and technologies. It starts with a brief history about cloud computing, tracing its roots. Then, architectural models of cloud services are described, and the most relevant products for Cloud Computing are briefly discussed along with a comprehensive literature review. The paper highlights and analyzes the most pertinent and practical network issues of relevance to the provision of high-assurance cloud services through the Internet, including security. Finally, trends and future research directions are also presented.

## 1. INTRODUCTION

The Cloud Computing (CC) market has been increasing very significantly. A recent study predicts that "from 2013 through 2016, $677 billion will be spent on cloud services worldwide" (Gartner 2013). The CC paradigm involves moving both data storage and applications into the network, offering to the users a ubiquitous access (Panagiotakis, et al. 2015) (Fernando, Loke and Rahayu 2013). These resources are available via the cloud in the same way as they would have been previously using local computers. Nevertheless, CC resources are made available via distributed virtual servers. Such virtual servers can be moved among distinct physical servers and dynamically adjusted in terms of their memory, CPU, or storage capacity, elastically following the users' load demand and satisfying their traffic requirements. CC is broadly accepted across the globe: diverse mobile operators (AT&T 2012) (BT 2014) (PT 2014) (DT 2014) (ND 2014) (MT 2014) and technological enterprises (Salesforce 2014) (Google_a 2014) (Microsoft_a 2014) (Amazon 2013) (Dropbox 2014) (Microsoft_b 2014) (Google_b 2014) are providing cloud services based on their network and computing infrastructures. In addition, four standardization organizations (one American: ANSI, which shares its CC vision with National Institute of Standards and Technology (NIST); and three European: CEN, CENELEC and ETSI) recently started a common initiative (ETSI 2013) in several relevant areas, namely electric vehicles, smart grids, machine-to-machine (M2M) communication, smart cities, and more pertinently to this paper, CC. Furthermore, recent literature describes cloud systems and comprehensively discusses the most relevant decision aspects to make the move to CC (Badger, et al. 2012) (Jamshidi, Ahmad, & Pahl 2013).

The performance of CC depends heavily on networking and, therefore, any limitations or failures of the networking infrastructure (e.g. inside and between data center domains) can seriously impair the support of data-intensive and/or high-performance cloud applications. Consequently, the deployment of CC solutions in distributed data centers, concurrently with the universal users' access to the Internet, is challenging the research and standardization communities to modify existing network functionalities. The need for these network changes is fuelled by emerging CC usage scenarios with dynamic load, data mobility, addressing/routing based on data alternatively to IP destination, heterogeneous resources, federation, and energy-efficiency. The article we present here aims to add to the literature a comprehensive and contemporary CC survey from the networking perspective. Various and extensive work on CC can be found in the literature, as shown in Table I.

Table I. Cloud Computing Surveyed Contributions

| Number | Reference | Main Contribution |
| --- | --- | --- |
| 1 | (Vogels 2008) | Seminal work presenting important CC aspects from a hardware point of view: i) illusion of infinite computing resources available on-demand; ii) elasticity on resource usage according the demand; iii) pay per use of computing resources on a short-term basis |
| 2 | (Mei, Chan and Tse 2008) | Presents a qualitative comparison between cloud, service and pervasive computing paradigms; this comparison was based on the classic model of computer architecture: I/O, storage, and computation |
| 3 | (Buyya, et al. 2009) | CC is envisioned as a paradigm that could deliver computing as the 5th utility (after water, electricity, gas, and telephony) |
| 4 | (Armbrust, et al. 2009), (Armbrust, et al. 2010) | Discussions about top 10 obstacles to and opportunities for growth of CC |
| 5 | (Oracle 2010) | This paper presents an introduction to CC (i.e. essential characteristics; service and deployment models); it also discusses some cloud benefits and challenges |
| 6 | (Zhang, Cheng and Boutaba 2010) | Extensive state-of-the-art implementation of CC; comparison of representative commercial products; discussion around research challenges |
| 7 | (Duan, Yan and Vasilakos 2012) | A comprehensive discussion on Service-Oriented Architectures towards the convergence of Networking and CC |
| 8 | (Alamri, et al. 2013) | Focused on Sensor-Clouds |
| 9 | (Dinh, et al. 2013), (Fernando, Loke and Rahayu 2013) | Discussions on mobile cloud computing |
| 10 | (Fernandes, et al. 2014), (Subashini and Kavitha 2011) | Stresses the distinct security requirements imposed by distinct cloud service models |
| 11 | (Ali, Khan and Vasilakos 2015) | Discusses security vulnerabilities in mobile cloud computing |

The novelty of the work we present here, in relation to other surveys, is to discuss how the network architecture, protocols and algorithms should evolve to support cloud services more capably in highly dynamic and resource-constrained environments. In this way, we discuss aspects related to the future evolution of cloud systems, namely: reliable and efficient allocation of networking resources including virtualization and emergent security aspects. Another value of the current paper is to provide a single source of information compiling all the relevant CC studies, and providing readers with a concise update of this area.

The paper is also well aligned with the emergent networking proposals from both academia and standardization bodies to meet new cloud requirements. The authors have made an effort to assemble cloud resources and references and to present them at two levels; first, for those readers who are seeking to build knowledge on this topic; and second, for those seeking to progress their research. Finally, we identify current open issues that may form a barrier to the successful deployment and management of cloud services in future networks.

*Organization of the Paper*

The main aim of this article is to review and analyse the major network functionalities that need to be modified or tuned to support the emergent properties of CC, using the present Internet infrastructure as a foundation. This contribution is organized as follows. Section 2 presents the main CC fundamentals and concepts, as well as tools and technologies to build clouds; this can help a non-specialized CC reader throughout the paper. Then, in section 3, we narrow our discussion with a comprehensive review of recent literature discussing challenges imposed by CC in the current networking infrastructures. To structure our discussion a list of relevant networking aspects is suggested. Section 4 outlines research directions for future networks in support of CC applications or services; this discussion is driven by representative scenarios, namely the Internet of Things (IoT) and Network Functions Virtualization (NFV). In section 5, we discuss current and future security challenges for cloud systems. Section 6 summarizes selected research challenges that could be addressed as future work. Finally, section 7 concludes the article.

The next section offers background information, mainly dedicated to readers who are building their knowledge in CC. Readers already specialized in CC could jump to section 3.

## 2. BACKGROUND OF CLOUD COMPUTING

This section introduces the main fundamentals and concepts that may be needed to follow the paper. We briefly present the historical evolution of CC; then we discuss the foundational technologies of CC, and compare the different CC service models.

### 2.1 History and Emergence of Cloud Computing

This section presents the most relevant aspects related to the history of CC. We start, however, with the origin of "cloud"; this word means an abstraction of the underlying infrastructure (computers, networks, data storage) that enables the normal operation of any CC system. It is also why network infrastructures have for many years been represented by an iconized "cloud", hiding its complex details from non-specialized individuals. The additional words presented together with "cloud" identify the scope of that "cloud", and it could be for example any of the following: computing, networking, mobile computing, and sensor networks. In addition, CC glossaries are available in (CCGa 2014) (CCGb 2014). Furthermore, some CC taxonomies are in (Rimal, Choi and Lumb 2009) (Beloglazov, et al. 2011). Table II briefly shows the historical evolution of CC since the 1960s until 2011.

More recently, in 2013, an international congress (Services 2013) gave special attention to Big Data Research and its major impact on social development (Obama 2012). Big Data is a recent trend (Ward and Barker 2013) (Diebold 2012) (Press 2013) which aims to extract pertinent knowledge from large-scale, complex, and unstructured data. This work is being carried out by numerous organizations including NSF, DoD, and DARPA. Some DARPA Big Data projects related to CC are described in (DARPA_a 2013) (DARPA_b 2013). Big Data implementation strongly depends on the existence of Internet cloud solutions to support big data storage, to scale up the distributed/parallel processing power, to enhance collaborative work, and to support the efficient, secure, and private access of mobile terminals to heterogeneous data and services (Moura and Serrão 2015).

Table II. Cloud Computing Historical Evolution from 1960s until 2011

| Organization / Project | CC Related Main Achievement | Year(s) |
|---|---|---|
| IBM | Mainframe time-sharing technology | 1960's |
| MicronPC (changed to Web.com) | Initial provider of websites and web services to small businesses and consumers | 1995 |
| Salesforce | Enterprise-level applications to which end users could have access via their Internet connections | 1999 |
| Amazon | Mechanical Turk was offered as an online marketplace for work | 2002 |
| Amazon | The first widely accessible CC infrastructure service (Elastic Compute Cloud - EC2). | 2006 |
| Academic Cloud Computing Initiative (ACCI) project | The ultimate goal of this project was to prepare students to explore the new potential cloud systems could offer at that time | 2007 |
| Google | Google Docs avoided the need for end-users to have locally licensed and always updated applications in their devices because the applications were stored in a remote and centralized location; collaborative working was in this way much easier to deploy | 2007 |
| Eucalyptus, OpenNebula | These were launched as the first open-source computing toolkits for managing clouds | 2008 |
| Microsoft | Windows Azure was launched a cloud solution | 2010 |
| IBM | The Smarter Computing framework was announced including CC as a relevant tool | 2011 |

Clearly, CC evolution is currently related to the increasing popularity of Big Data. In fact, CC provides the necessary computation, storage, applications, and networking, which support Big Data applications. These applications empowered by CC solutions can extract very useful information to guide better decisions in many usage areas like business, finance, politics, education, military, industry, transportation, research, and even healthcare (Griebel, et al. 2015) .

There are also important research areas for Future Networks with a strong relation to CC. These include Internet of Services, Grids, Service Oriented Architectures, Internet of Things (IoT), and Network Functions Virtualization (NFV). These two last areas (i.e. IoT and NFV) are discussed at the end of the paper in terms of network challenges that should be addressed to satisfy their major requirements when they are implemented within the cloud.

In the next sub-sections, the concepts and technologies of CC are discussed.

**2.2 Definition of Cloud Computing**

There is an analogy between electricity and CC. Electricity is, of course, a utility where we expect a certain set of qualities (e.g. always-available, "five nines" reliability) and we believe that CC should aspire to be a utility too (Voorsluys, Broberg and Buyya 2011).

CC refers to computing services that are provided within a cloud infrastructure and accessed on demand by customers, so that the customers do not have to be concerned with the details of service provisioning.

Now, we present some definitions of CC. (Buyya, et al. 2009) have characterized it as follows: "Cloud is a parallel and distributed computing system consisting of a collection of inter-connected and virtualized computers that are dynamically provisioned and presented as one or more unified computing resources based on service-level agreements (SLA) established through negotiation between the service provider and consumers." The National Institute of Standards and Technology (NIST) (Mell and Grance 2011) has defined CC as "… a model for enabling convenient, on-demand network access to a shared pool of configurable computing resources (e.g., networks, servers, storage, applications, and services) that can be rapidly provisioned and released with minimal management effort or service provider interaction." Further definitions about CC are available in (Voorsluys, Broberg and Buyya 2011).

In recent years, the rise of CC is due to several foundational technologies that are discussed in the next sub-section.

**2.3 Foundations of Cloud Computing**

CC resulted from the convergence of several technologies belonging to four distinct fields: hardware (e.g. virtualization), distributed computing (e.g. grid computing), the Internet (notably service-oriented applications), and network management (Voorsluys, Broberg and Buyya 2011).

Cloud services are normally situated in data centers each deploying thousands of computers. These systems need to scale up to very high rates of service demand with an acceptable processing time, and also with low costs in terms of energy and hardware. To achieve these goals, a conceptual cloud model such as the one shown in Figure 1 could be adopted.

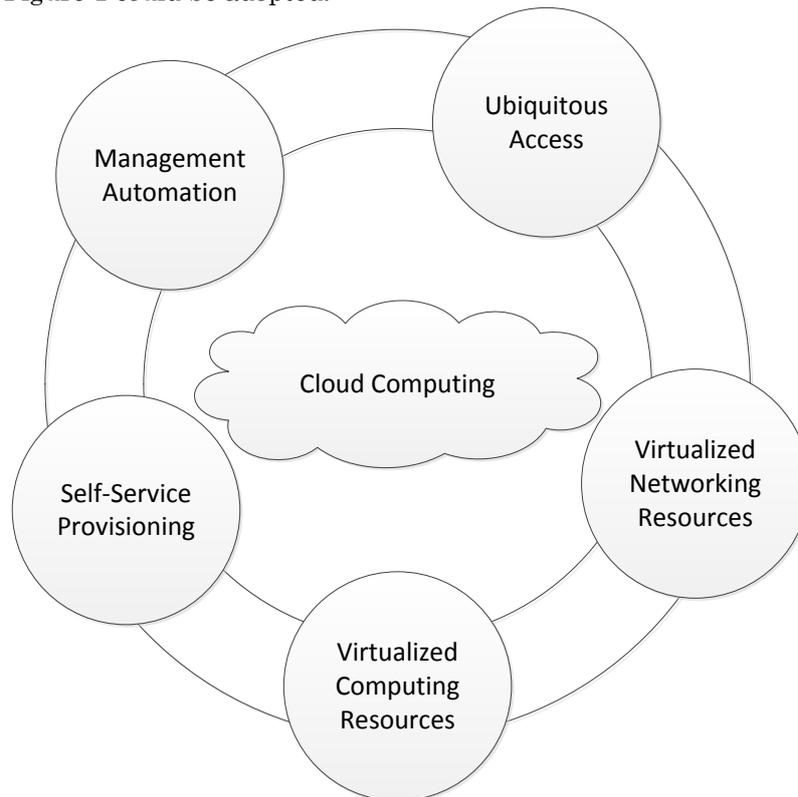

Fig.1. A common view of architectural foundation elements of Cloud Computing.

In the model of Figure 1, the virtualization of computing resources can offer significant advances in the following aspects: security, reliability, compatibility, utilization, maintenance, load balancing, and problem recovery. In this way, a virtualization platform normally requires a Virtual Machine Monitor (Hypervisor), which could run directly above the hardware resources of a physical computing machine (host) and immediately below the virtual machines (guests). There are many virtualization platforms underlying CC, as discussed in (Voorsluys, Broberg and Buyya 2011). There are three types of Hypervisor depending in what layer the Hypervisor entity is running. The first is designated as Type 1, and groups all the virtualization platforms with their Hypervisor running directly above the hardware of the host machine. The second, Type 2, represents all the virtualization platforms with their Hypervisor running directly above the operating system of the host. Finally, Type 3 is designated as hybrid and classifies all the virtualization platforms in which the Hypervisor runs at the same layer as the host operating system.

An example of a virtualized computing resource is that of grid computing, whose main goal is to distribute the processing of high complexity and/or time-consuming applications across a group of distinct machines to obtain the intended application results as fast as possible. Grid computing is very relevant in some specific use cases such as drug design, climate modelling, protein analysis, and physics research. GridGain (GridGain 2014) is an open cloud platform to develop and run Java applications. It can split an initially complex task into multiple subtasks using the MapReduce programming model (Jin, et al. 2011) (Li, et al. 2014). These subtasks are delivered to distinct machines and each one of these subtasks is executed in parallel. At the final stage, the processing results of all the subtasks are aggregated (i.e. reduced) back to one final result. An issue associated with some grid systems is the portability barrier imposed by the diverse operating systems, libraries, compilers and runtime environments available in the computing machines forming the grid processing environment. To overcome these issues, virtualization has been identified as a potential solution (Keahey, et al. 2005).

Returning again to Figure 1, alongside the architectural element of virtualized computing resources, a CC system also requires virtualized networking resources, ubiquitous (i.e. reliable / efficient / secure) access, self-service provisioning, and management automation. As these elements are self-explanatory, we refrain from discussing them in this section, with the exception of management automation. In fact, the high complexity associated with CC systems has motivated the research on management automation. This aims to automatically optimize resources usage and adapt in real time to the customers' needs and operational system status (Murphy, et al. 2010). As large data centers from CC providers have highly dynamic demands and workloads, these must be managed in an efficient way (Kim and Parashar 2011). In the subsequent subsection, we discuss some important architectural aspects of CC systems such as the diverse service models.

**2.4 Cloud Computing Service Models**

We next discuss CC services, depending on the degree of awareness that cloud providers give to subscribers to control the supplied services. Each one of the following sections discusses a single CC service model. In the beginning of each section we highlight the differences between the associated model and other possible CC models concerning how the control scope is divided among the cloud provider and clients. Then, some real deployments of that model are presented. Finally, the strong and weak functional aspects of each model are also discussed.

*Software as a Service (SaaS)*

A Software as a Service (SaaS) cloud system allows customers to have access to applications and settings that have been deployed by the provider. The clients can have access to these cloud applications using a simple browser. In SaaS, the software stack is controlled in its vast majority by the cloud provider and, significantly, the cloud subscriber is only authorized to control the application level. For example, the subscriber cannot configure the middleware or even the operating system of each virtual machine. Figure 2 illustrates how the cloud provider and subscribers share among them the control and management responsibilities through a vertical software stack comprising distinct layers.

A SaaS cloud has the potential to join and compose services from distinct providers. In this way, composed elements can provide high-value solutions for use cases where a single element does not fulfil all the requirements. Many SaaS proposals are now offered. For example, the Salesforce platform offering diverse software components can build innovative, collaborative, community, secure, personalized, mobile and real-time applications for customers (Salesforce 2014). Similarly, the Programmable Web offers a diverse and numerous set of Application Programming Interfaces (APIs) (ProgrammableWeb 2014). The Programmable Web API lets a customer to find, retrieve and interconnect APIs, mashups, member profiles and other content from the Programmable Web repository, producing a variety of interesting, novel, customized and on-the-fly services from finding specific product retailers to weather forecasts or geographical maps, although sometimes in a rather limited way.

The main advantage offered by SaaS cloud systems is that it almost eliminates the deployment and maintenance tasks for a customer, who can then rely on the SaaS provider to carry these out instead (Sridhar_a 2009).

The SaaS products, in spite of their simplicity to offer pre-defined applications that can be settled together in innovative designs, have some drawbacks. As shown in Figure 2, the cloud subscriber cannot add a new application to the portfolio of the SaaS provider. In fact, the cloud subscriber only has a limited access to personalize any required application. Other limitations imposed to the cloud subscribers by the SaaS provider include the fact that only the SaaS provider can monitor the application-delivery performance (i.e. configure the resources allocated to each client). In this way, cloud subscribers cannot in any way scale up or down the allocated resources according the storage needs or the data traffic changes overtime simply because they cannot configure the middleware (Figure 2). To satisfy all these requirements that are not ensured by SaaS products, Platform as a Service (PaaS) solutions can be a good alternative option, as explained in the next subsection.

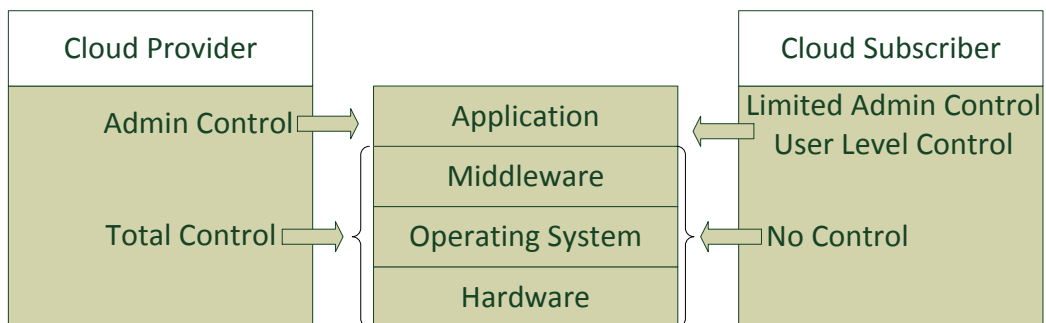

Fig. 2. SaaS provider/subscriber control responsibilities.

*Platform as a Service (PaaS)*

To allow customers full control of applications and configurations according to their particular requirements, a PaaS solution (see Figure 3) can be used alternatively to SaaS solutions. In fact, comparing these two service models, shown in Figures 2 and 3, a PaaS provider offers its customers an additional Application Programming Interface (API) for dynamically adjusting the computational resources (e.g. memory, storage disk) according to customers' requirements. Some very popular PaaS offerings are available in (Google_a 2014) (Microsoft_a 2014).

The platforms offered by PaaS vendors force their applications to be coded in a specific language, following their own API. This creates huge difficulties to move legacy applications to a new PaaS environment or to move applications between distinct cloud providers. This last scenario occurs if a customer takes the decision of changing its cloud provider. These problems could be avoided if the PaaS vendors agree on a standard API or if the customers decide to subscribe with an IaaS cloud provider (see next subsection).

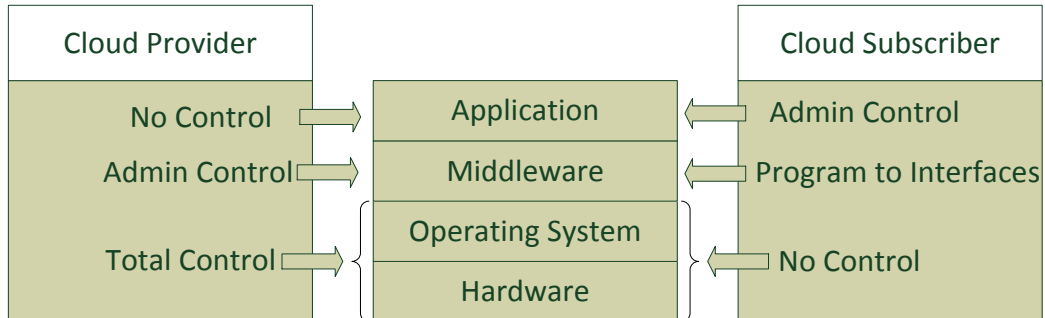

Fig. 3. PaaS Provider/Subscriber Control Responsibilities.

*Infrastructure as a Service (IaaS)*

Where a cloud provider allows its subscribers to have total control of virtual machines (i.e. a customer can choose the operating system for each dedicated virtual machine), we have an Infrastructure as a Service (IaaS) cloud model. Figure 4 illustrates how the cloud provider and cloud subscribers share among them control and management responsibilities when an IaaS model is being used. An IaaS cloud system provides its customers with several fundamental technological resources such as processing, storage and networking. In this way, the customers can install and run distinct software and services of their own choice but without access to or management of the underlying physical system, as shown in Figure 4, though possibly with a limited authorization to set up some networking elements (e.g. firewalls, NATs).

In the present case, the virtualization should be used to guarantee to each cloud subscriber a machine with a full operating system that is completely independent from the remaining operating systems associated with other subscribers, in spite of all these operating systems running over the same hardware. Figure 4 illustrates, just above the hardware, the layer designated by the Virtual Machine Monitor (VMM), or commonly the 'hypervisor'. The hypervisor uses the same hardware and shares its computational resources among diverse Virtual Machines (VMs). Each VM operates like a real machine but is completely isolated from the remaining VMs. In this case, the VM appears to the subscriber like a standalone machine that can be completely configured by that subscriber in various aspects, namely: i) switch on/off the VM; ii) install any supported guest operating system, iii) install a full set of preferred applications/services; iv) adjust computational resources such as memory, CPU cores, data storage or network interfaces. The previous VM configuration can be easily made through remote command messages sent to the provider's cloud.

Amazon EC2 is an IaaS cloud model that enables developers to build applications that are resilient against failure situations (Amazon 2013). This is a major advantage over the PaaS cloud model discussed in the previous section. Amazon EC2 offers a very flexible virtual computing environment. In fact, this model allows its customers, using a simple web browser interface, to configure, not only, the diverse VM operational aspects referred in the previous paragraph but also specifying the correct number of VMs to properly satisfy the customer demand or requisites. We next discuss some relevant Amazon EC2 features, such as: i) Elastic Block Store (EBS), ii) cloudwatch, iii) auto scaling, iv) elastic load balancing, v) High Performance Computing (HPC), and vi) VM import/export.

First, the Amazon EBS feature provides storage network volumes that can be attached in a reliable and elastic way to already running Amazon EC2 instances. Second, the cloudwatch feature monitors Amazon Web Services (AWS) resources and performance parameters generated by customers' applications, enabling the automatic tuning of virtual resources according to the customers' needs. The third Amazon feature designated by auto scaling tunes automatically the Amazon EC2 capacity according to the processing load. With auto scaling, it is possible to adjust the number of Amazon EC2 instances being used, according the demand and minimizing the cost. Fourth, using the elastic load balancing feature, it is possible to automatically distribute incoming application traffic among several Amazon EC2 instances. It also enables a system with high reliability, detecting unhealthy EC2 instances and automatically rerouting the data traffic destined for these to alternative healthy EC2 instances. Using the fifth Amazon feature designated by High Performance Computing (HPC), the AWS customers are able to solve complex scientific and/or engineering problems exploring the potentialities of distributed applications that assist their research/work in physics, chemistry, biology, engineering, or computer science. The last EC2 feature, i.e. the VM import/export, enables a customer to easily import previously configured Amazon EC2 instances as ready-to-use machines and afterwards export these back to the customer virtualization infrastructure. It allows the customer to deploy workloads across his IT infrastructure with a controlled cost and always satisfying customer requirements including security, configuration management, and compliance.

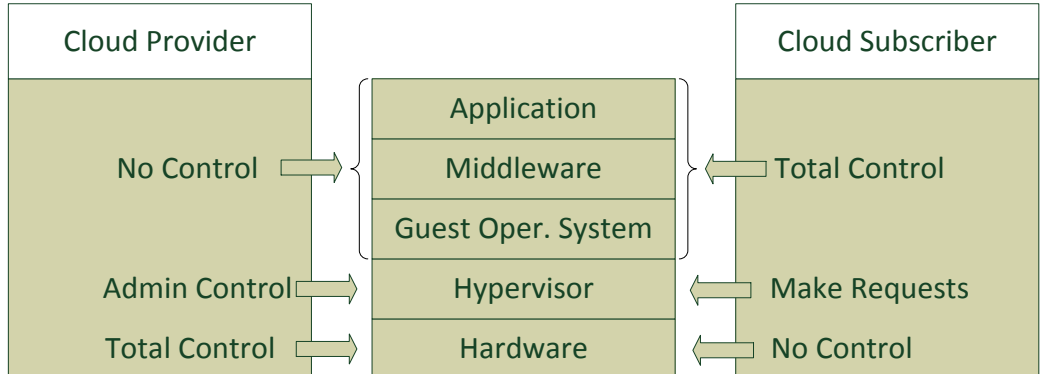

Fig. 4. PaaS Provider/Subscriber Control Responsibilities.

Another current IaaS solution is available (Eucalyptus 2014). In addition, there are some well identified and challenging issues to be solved in IaaS cloud services, such as virtual networking, cloud extension, and cloud federation (Azodolmolky, Wieder and Yahyapour 2013). In addition, the programmability through a simplified API has been proposed very recently as a new concept to manage an IaaS cloud infrastructure. In fact, through a simplified API, the applications in an on-demand way can control distinct cloud system aspects, such as resource allocation (Wickboldt, et al. 2014) and creation and management of overlay networks (Strijkers, et al. 2014). As a proof of concept, it has been shown how to create an IPv6 network over a number of cloud locations around the world (Boutaba, et al. 2014) (Strijkers, et al. 2014).

The reader should note that the previous cloud service taxonomy formed by only three options (i.e. SaaS, PaaS, and IaaS) is rapidly extending to "X as a Service", where X namely includes Backend, Business Process, Database,

Information, Infrastructure, Storage, Platform, Security, Software, Network, and more generically Everything.

Although the functionality of cloud technologies has been comprehensively investigated, less attention has been devoted to relevant aspects of networking that can significantly impair the performance of cloud systems. This novel perspective is studied in the next section.

## 3. NETWORK ARCHITECTURE IN CLOUD COMPUTING

This section provides a comprehensive and structured review of recent literature, including relevant standardization contributions. The structure of this section (and Section 4) is strongly related with Figure 1 (i.e. CC Architectural Elements), as Table III shows.

Table III. CC Architectural Element from Figure 1 vs. Discussion Topics Covered by the Current Survey

| CC Architectural Element from Fig. 1 | Discussion Topics Covered by the Current Survey |
| --- | --- |
| Ubiquitous Access | Reliable, Efficient, and Secure Communications |
| Virtualized Computing Resources | Not Covered |
| Virtualized Networking Resources | Virtual Networking |
| Management Automation | Other Aspects (Elasticity, Federation, Interoperability, Cooperation, MCC, NFV, Inter-Cloud Architectures, IoT) |

In the following subsections we discuss the most significant research / standardization efforts inside the networking area mainly for supporting, with enhanced performance, some CC emerging applications. These applications are setting new system requirements such as elastic load, dynamic allocation of network resources, and secure services distributed between private and public infrastructures. Table IV shows more details about the organization of our subsequent discussion through section 3 on short-term networking challenges.

Table IV. High-Level Structured List of Networking Research Activities in relation to Cloud Computing

| Main Area | Goal |
| --- | --- |
| Reliable communications (Sub-section 3.1) | Support protocol heterogeneity; control of flow rate to avoid either receiver's buffer overflow or congestion; simplify the synchronization between transmitter and receiver; error detection |
| Efficient communications (Sub-section 3.2) | Balance the load among alternative paths; higher data rates; provision for larger frame sizes; multiple virtual channels operating in parallel over a single physical channel; some extensions to Open Shortest Path First (OSPF); some BGP enhancements |
| Virtual networking (Sub-section 3.3) | Virtual switch management; VLANs management with VM migration; SDN |
| Other important areas (Sub-section 3.4) | Elastic allocation of cloud resources according the load variation; cloud federation; interoperability |

### 3.1 Reliable Communications

Full communications reliability should be supported to deliver data messages to the intended recipient(s) within a cloud infrastructure in a correct and timely way. Currently, standardization bodies are working towards several solutions to enhance communications through the available and forthcoming networking infrastructures. In the following text, we discuss the more relevant enhancements proposed within both IEEE and IETF that can be applied into cloud networking environments. The first enhancement implies a reliable link protocol, i.e. Fiber Channel over Ethernet - FCoE (ANSI/INCITS 2009), which is an encapsulation mechanism. It can be used to simplify and enhance the interconnection between a classical Ethernet network and a distributed storage area network (SAN). This encapsulation requires some changes in the Ethernet operation namely, the usage of an additional mapping between Fibre Channel N_port IDs (i.e. FCIDs) and Ethernet MAC addresses.

To avoid losing frames, a second enhancement is appropriate, IEEE 802.1Qbb (IEEE_a 2011), which uses a priority-flow control mechanism to selectively counteract losses due to the receiver's buffer overflow. This mechanism uses a pause control message that is sent by a receiver to the sender after the former predicts the potential for buffer overflow. Upon receiving a PAUSE frame, the sender responds by stopping transmission of any new frames through the link interconnecting both of them until the receiver is ready to accept frames again. The novelty of this solution is that it can control the transmission of flows in different ways depending on the diverse flow types.

The third enhancement proposal, IEEE 802.1Qau (IEEE_b 2010), also avoids transmission losses as in the previous proposal but now avoiding losses that can occur during frame transmission, for example due to switch buffer overflow. This proposal supports congestion management of long-lived data flows within network domains of limited bandwidth-delay product. This is achieved by enabling switches to signal congestion to end stations capable of transmission rate limiting to avoid frame loss. There is also a major difference between the current controlling mechanism and the second mechanism, 802.1Qbb. The latter is a hop-by-hop mechanism, whereas the former, 802.1Qau, operates end-to-end.

The fourth proposal, IEEE 802.1Qaz (IEEE_c 2011), defines enhancements to transmission selection to support allocation of bandwidth amongst traffic classes, plus a protocol for controlling the application of Data Center Bridging features.

The last proposal is associated with the IETF ConEx working group that is chartered to work on a congestion exposure mechanism for IPv6 networks (Mathis and Briscoe 2014). This mechanism allows data sources to notify the network about the congestion suffered by previous packets of the same data flow. A very recent contribution to this working group (Briscoe and Sridharan 2014) shows how to police congestion at data center ingress nodes and thereby how traffic shaping can be applied to provide suitable per-flow performance. This functionality based on a feedback congestion mechanism to the ingress nodes avoids the configuration of any of the internal data center network switches with flow related configuration (Briscoe and Sridharan 2014). Another way of solving the congestion problem in clouds is based on OpenFlow (McKeown, et al. 2008), which uses a centralized design with controllers and some flow configuration in the switches establishing the data path.

### 3.2 Efficient Communications

Some ongoing work is investigating efficient communications, which is obviously very important in a cloud scenario. The reader may note some overlap between what will be discussed now and what was already discussed in the last section. The initial case we discuss here is Shortest Path Bridging (SPB), specified in the IEEE 802.1aq standard (IEEE_d 2012). It is a computer networking technology intended to simplify the creation and configuration of networks, while enabling multipath frame forwarding. SPB is the replacement for the legacy spanning tree protocol (STP) (IEEE_f 2004). Applying legacy STP into the typical flat (non-hierarchical) topology of current data centers is not recommended because it would force the existence of a root bridge and a hierarchical tree of switches without loops, which potentially create non-optimum switching paths inside the local network of the data center, and consequently cause the following problems: non-balanced load among the local links, some links become highly congested, and a significant delay growth of frame transmission among servers, which can negatively impact the overall datacenter performance. Alternatively, SPB allows all paths to be active with multiple (and eventually equal) cost paths, and provides much larger layer 2 topologies (i.e. up to 16 million virtual local area networks (VLANs) compared to the traditional limit of 4,096). It also supports faster convergence times, and improves the efficiency of the mesh topologies through increased bandwidth and redundancy, allowing traffic to be balanced within a mesh across all possible paths. Additional related standardization work is being carried out by IETF in Transparent Interconnect of Lots of Links - TRILL (Eastlake, et al. 2011) (Perlman and Eastlake 2011). The main goal of TRILL is to encapsulate each Ethernet frame within another envelope (i.e. using the outer TRILL header), which acts like a layer 3 envelope, and then this encapsulated frame can be routed using all the Layer 3 routing techniques that have evolved over the years, including shortest paths and multipath techniques (Perlman and Eastlake 2011). It allows a fairly large Layer 2 cloud to be created, with a flat address space, so that nodes can move within the cloud without changing their IP addresses. The cost is a small overhead induced by the outer header added to each frame traversing the cloud infrastructure. At the time of writing, there is intensive activity at IETF on this topic. However, with the exception of (Amamou, Haddadou and Pujolle 2014), we could not find any recent other work in this field. Therefore, it seems clear that further research would be valuable.

Another interesting enhancement aspect involves the Gigabit Ethernet, from 1 to 100Gbps and beyond, which is very well covered in (Stallings, 2015). This describes some enhancements to the MAC layer, such as provision for larger frame sizes; the usage of 2 control bits beyond the data bits to enable both easier transmitter/receiver synchronization and error-detection; and finally a multi-lane distribution allowing a single physical link to work as multiple parallel channels.

The design of networking systems that enable communications among data centers can involve the utilization of enhanced versions of high-layered routing protocols such as OSPF (Retana, 2013). In this case, OSPF is used internally in each data center to support the routing of packets based on their final IP destination address through a path with the lowest cost. Following this track, (Retana, 2013) discusses three extensions to the Open Shortest Path First (OSPF) protocol that have direct applicability to efficient and scalable network operation in highly meshed environments, typically the ones present in the data centers. Specifically, the application extensions to OSPF to reduce flooding in Mobile Ad Hoc Networks (MANET), demand circuits designed to support on-demand links in wide-area networks, and OSPF stub router advertisements designed to support large-scale hub and spoke networks are considered in a typical data center network design; these sorts of protocol improvements could affect the scaling of data center environments. On the other hand, the Border Gateway Protocol (BGP) can be used in the core communications among data centers. In this case, BGP is responsible for setting up Multiprotocol Label Switching (MPLS) that forwards traffic through the core network based on short path vector labels rather than long network prefixes. Nevertheless, BGP needs to be enhanced to support QoS/QoE per flow. To achieve this, SDN (Astuto, et al. 2014) can potentially be very useful (Gupta, et al. 2014). Another way in which BGP can be enhanced is the availability of multiple routes to a given destination, where each of the routes has a different "exit point" from the local Autonomous System (AS) (Mohapatra, et al. 2015). If this enhancement will be deployed in a communications scenario among data centers, due to the presence of multiple paths, the following benefits can be attained: reduce the restoration time after a failure, enable load balancing of traffic, help contain the failure to the local AS where the failure occurs, and allow one to bring down a router for maintenance without causing significant traffic loss among the data centers due to the availability of alternate exit points from the AS to a given destination.

### 3.3 Virtual Networking

A typical physical host of a (cloud) data center has a hypervisor that enables diverse Virtual Machines (VMs, or guests) to run over the same host hardware. In order to offer a stronger interworking and interoperability between system and network elements, the virtualization of networking resources within a cloud infrastructure is becoming a very important requirement. In this way, the hypervisor is also associated with a virtual switch (VS). This device switches Layer 2 traffic among the VMs running in the same physical server. The VS learns about MAC addresses in a different way from a traditional switch because the former assumes by default that all frames with an unknown destination MAC should be forwarded over the uplink to the physical external switch. This default behavior can potentially create some security threats such as packet sniffing and spoofing (Wu, et al. 2010). In addition, the VS can switch traffic among the intra-machine VMs according to pre-defined policies, which can control broadcast and Virtual LAN (VLAN) traffic. As an example, a VS access control list for security reasons can disallow two VMs located on the same physical host to have a direct communication between them. In this way, the VS is like a software routine that controls the traffic features of aggregation and access control associated to its virtual ports within a physical server containing diverse VMs (Sridhar_b 2009) (Chowdhury and Boutaba 2010).

VSs also have some disadvantages (Sridhar_b 2009). They can potentially create serious practical problems with traditional network architectures (Layland 2010). One problem is related to the configuration of VLANs: each time a VM moves to other physical server, it is necessary to reconfigure the VLAN through distinct switches. This coordination among the switches could be complex, with a large latency, and sometimes impossible due to the fact these switches are from distinct vendors, each one with its own proprietary firmware and incompatible with the others. A potential solution to these problems is separating the control functions from the network switch and placing them in accessible control servers. This separation can be supported by a Software Defined Networking (SDN) architecture (Astuto, et al. 2014) and using the OpenFlow protocol (Stallings 2013). Some disadvantages of these new proposals include: the LAN overhead due to extra signaling/control traffic; the lack of robustness against the failure of a control server (i.e. the typical problem of a centralized solution); and the VLAN reconfiguration in aggregation and core switches. Some possible solutions to these disadvantages are, respectively: optimizing the OpenFlow protocol to reduce its header size and reduce the number of signaling/control messages; the deployment of redundant SDN controllers with horizontal/vertical communication among them; and the deployment of hairpin switching, which is a new approach to allow the visibility of intra-VM traffic to external network switches. Hairpin switching is being actively discussed inside the IEEE (IEEE_e 2012). The IEEE proposes a tagging approach in the header frame to perform hairpin switching.

SDN can be used to enhance some important aspects of clouds such as network virtualization and security, as illustrated in Table V.

Table V. SDN Proposals to Support Network Virtualization and Security in Cloud Systems

| Topics | Network Virtualization | Security |
|---|---|---|
| SDN | (Jain and Paul 2013), (Corradi, et al. 2014), (Vestin, et al. 2013), (Banikazemi, et al. 2013) , (Benson, et al. 2011), (Chowdhury and Boutaba 2010) | (Shin and Gu 2012) |

*Network Virtualization*

A SDN solution that dynamically manages networking tunnels is discussed in (Jain and Paul 2013). They proposed a proactive solution to deploy overlay tunnels. These tunnels could end either at virtual switches controlled by hypervisors or physical switches. This hybrid design is very common in data centers where also several tunneling technologies are used (Jain and Paul 2013).

OpenStack controls large pools of compute, storage, and networking resources belonging to private/public clouds. This software enables the system management through a dashboard or via the OpenStack API. OpenStack works with popular enterprise and open source technologies, making it ideal for heterogeneous infrastructures (Corradi, et al. 2014).

The current distributed control plane of wireless networks is suboptimal for managing the limited spectrum, allocating radio resources, implementing handover mechanisms, managing interference, and performing efficient load-balancing between cells. SDN-based approaches represent an opportunity for making it easier to deploy and manage different types of wireless networks, such as WLANs and cellular networks, eventually supporting traffic offloading. Traditionally hard-to-implement but desired features are indeed becoming a reality with the SDN-based wireless networks. These include seamless mobility through creation of on-demand virtual access points (VAPs), downlink scheduling (e.g., an OpenFlow switch can do a rate shaping or time division), dynamic spectrum usage, and enhanced intercell interference coordination (Vestin, et al. 2013). Other centralized SDN controllers such as Meridian (Banikazemi, et al. 2013) can be used to manage target specific environments such as data centers, cloud infrastructures, and carrier grade networks.

SDN can also potentially offer networking primitives for cloud applications, solutions to predict network transfers of applications, mechanisms for fast reaction to operation problems, network-aware VM placement, QoS support, real-time network monitoring and problem detection, security policy enforcement services and mechanisms, and enable programmatic adaptation of transport protocols (Benson, et al. 2011). SDN can help infrastructure providers to expose more networking primitives to their customers, by allowing virtual network isolation, custom addressing, and the placement of middleboxes and virtual desktop cloud applications. Further information about network virtualization is available in (Chowdhury and Boutaba 2010).

*Security*

An already diverse set of security and dependability proposals is emerging in the context of SDNs. As an example, (Shin and Gu 2012) discusses a proposal to monitor cloud infrastructures for fine-grained security inspections. It automatically analyzes and detours suspected traffic to be further inspected by specialized network security appliances, such as deep packet inspection systems.

### 3.4 Elastic Allocation, Federation, and Interoperability

This sub-section deals with additional networking research activities in CC, which are summarized in Table VI.

Table VI. Summary of Other Networking-Based Research Activities in Cloud Computing

| Main topic within the area of cloud computing | Reference |
| --- | --- |
| Elastic allocation of cloud resources according the load variation | (Hu, et al. 2012), (Seibold, et al. 2012), (Birke, Chen and Smirni 2012) |
| Cloud federation | (Sridhar_b 2009), (DMTF_a 2014), (Sen, et al. 2013), (Sakamoto, et al. 2012) |
| Interoperability | (DMTF_a 2014), (OASIS 2014), (SNIA 2014) |

*Elastic Allocation of Cloud Resources According the Load Variation*

The authors of (Hu, et al. 2012) discuss solutions for addressing as well as routing and forwarding using layered network topologies. Also, according to (Seibold, et al. 2012), running large databases requires the use of virtualization in order to cope efficiently with peak demands. They propose a cooperative approach, in which the database management systems communicate their request for resources (typically then deployed by virtual machines) and adjust their resource usage. Additionally, a large-scale survey about workloads for data centers (Birke, Chen and Smirni 2012) may be a great help for a reliable future planning. Finally, the authors of (Metri, et al. 2012) have conducted extensive sets of experiments on data centers' energy efficiency and have identified the need for accurate load prediction and how to set up the necessary virtual machines to fulfil that load in a completely dynamic way.

*Cloud Federation*

Another relevant area is cloud federation. (Sridhar_b 2009) has defined cloud federation as follows: "Cloud federation manages consistency and access controls when two or more independent geographically distributed clouds share either authentication, files, computing resources, command and control, or access to storage resources."

Some of the most important features in cloud federation are, as discussed in (Sridhar_b 2009):

- A customer who considers multiple cloud services (e.g. SaaS) should instead use a single sign-on (SSO) scheme to authenticate that customer only once, irrespective of the cloud service providers involved. This requires a third-party authentication server that operates in a distributed way among customers and service providers. This authentication server initially receives the customer credentials and authenticates that customer. After this, the authentication server provides the credentials of the already authenticated customer to the selected cloud service provider. (Kerberos 2014) is a security/trust framework that can support previous functionality.
- All the computing and storage resources of a VM are normally saved in files. To support VM migration (Medina and Garcia 2014) transparently and reliably among distinct cloud technologies, it is necessary to use a portable format to save and share the complete status information among different technologies without any compatibility problems. In this way, the Desktop Management Task Force (DMTF) has produced a specification designated by Open Virtualization Format (OVF) to completely describe the VM in a neutral and universal format for use across many vendor platforms (DMTF_a 2014).
- Cloud federation is a very recent aspect in the cloud arena, fuelled by the user's need for pervasive access to the application's portfolio and data. Also, the application could be from a provider, and the data being used by that application can be stored in another provider. Assuming this type of emergent scenario, the providers will be much better off in terms of business if they cooperate. Therefore, the providers are likely to establish peering agreements, producing compatible APIs to offer easy access to their clouds. In fact, this could occur even before the standardization organizations produce any standards in this area. If this occurs, the provider and vendor innovation could significantly impact the successful implementation of cloud federation.
- The success of cloud federation implementation also depends on the coordination level between management and billing systems as well as the adoption of new business models (Sen, et al. 2013) for this new environment. In this way, the customers are expected to be billed according to the amount of resources/data they use from each provider's cloud. In addition, cloud service providers can adopt some mobile operator business models already being used to support peering/roaming agreements among them.

A recent piece of work proposes a way of sending requests about energy prices to (federated) data centers to help optimize the savings in electrical energy (Sakamoto, et al. 2012). They have developed a management policy to help target the requests to where electricity is cheaper. Their results suggest that reductions on electricity costs of 15% are possible.

*Interoperability*

There is active standardisation work in the CC area on interoperability. Some coordination efforts have been established to minimize the problems of redundancy and incompatibility among specifications.

The Desktop Management Task Force (DMTF) has specified OVF (Open Virtualization Format) by means of which the VM full configuration and status can be written into files, and eventually migrated among physical machines, using distinct hypervisors (DMTF_a 2014). Also, the DMTF's Open Cloud Standards Incubator is interested in studying the following aspects: cloud portability (working with multiple providers), federation of cloud providers, and service adaption to varying requirements.

There is also the Organization for the Advancement of Structured Information Standards (OASIS) which views Service-Oriented Architectures (SOAs) as a basis of CC; SOAs are of course very popular in IT environments (OASIS 2014). More particularly, they are investigating CC as follows:
- Moving on-premise applications to private or public clouds.
- Enhancing the interoperability of cloud applications and services.
- Managing, in real-time, authorizations enriched by data that informs where users are, what they are doing, and which devices they are using.
- Simplifying the querying and sharing of data across disparate applications, clouds, and mobile devices.
- Developing a set of functional elements and measurable criteria or qualities that should be present in clouds deployed by public administrations.

The Cloud Storage Initiative (CSI) within the Networking Industries Association (SNIA) works on cloud-storage-related issues (SNIA 2014). CSI is proposing personalized cloud storage. They have developed a new interface designated as the Cloud Data Management Interface (CDMI); this allows cloud customers to associate to their data some metadata that informs the cloud provider about relevant data services (e.g. data special requisites, backup, archive, encryption, authentication, and authorization).

### 4. FUTURE NETWORK TRENDS FOR CLOUD COMPUTING

The discussion on the networking issues presented in the previous section underlines that the cloud deployment through the Internet obliges investigators to revisit traditional networking concerns, such as reliable and efficient communications, virtualization, security, resource allocation, and interoperability, due to the use of multi tenancy over a pool of shared virtual resources, notably computing, storage, and networking.

Moreover, future trends in computer communications have often been debated. In particular, the following vision has been presented (Huston, 2012): "It is also evident that the pendulum of distribution and centralization of computing capability is swinging back, and the rise of the heavily hyped Cloud with its attendant collection of data centers and content distribution networks, and the simultaneous shrinking of the end device back to a terminal that allows the user to interact with views into a larger centrally managed data store held in this cloud, appears to be back in vogue once more". In the following sub-sections we discuss relevant open issues in networking that could more effectively support CC. If correctly addressed, they could support the above vision about the evolution of computer communications and either attenuate or mitigate the networking issues that confront CC.

Table VII shows more details about the organization of our subsequent discussion through section 4 on future network trends for CC.

Table VII. High-Level Structured List on Future Network Trends for Cloud Computing

| Main Area | Goal |
| --- | --- |
| Reliable communications (Sub-section 4.1) | Solve the tradeoff between resource allocation and fault tolerance in resource-constrained systems; enhancement of Gigabit Ethernet |
| Efficient communications (Sub-section 4.2) | Cloud resources among tenants are urged to be shared in a safe and efficient ways within a cloud federated system; investigate and standardize relevant metrics to assess performance and energy efficiency of cloud systems |
| Virtual networking (Sub-section 4.3) | SDN could help to study fully collaborative, peer-to-peer and pervasive web scenarios, where the client-server paradigm could become obsolete |
| Other important areas (Sub-section 4.4) | Cooperation in Cloud Computing; Mobile Cloud Computing and Network Functions Virtualization; inter-Cloud Computing architectures; Internet of Things |

### 4.1 Reliable Communications

A recent contribution (Bodik, et al. 2012) optimizes the tradeoff between resource allocation (Chowdhury and Boutaba 2010) (Shieh, et al. 2011) (Ballani, et al. 2011) (Duffield, et al. 1999) (Ricci, Alfeld and Lepreau 2003) and fault tolerance (i.e. availability) (Agarwal, et al. 2010) (Amiri, et al. 2000) (Bansal, et al. 2008) (Yu, Gibbons and Nath 2006) in future resource-constrained systems. Meanwhile, the current growth in demand is accelerating the investigation into enhancements of Gigabit Ethernet to produce a 400 Gbps Ethernet standard. Looking beyond this milestone, there is a widespread consensus that a 1 Tbps will eventually be produced (Stallings, 2015).

### 4.2 Efficient Communications

Sharing computational, storage, and networking resources among cloud systems has been suggested in (Popa, et al. 2011). In addition, IETF work on Congestion Exposure (ConEx) (Mathis and Briscoe 2014) proposes a method for achieving congestion proportionality. However, this approach is still an open issue (Popa, et al. 2011). Sharing cloud resources in a conservative way, meaning that the unused cloud resources are shared in a safe and efficient ways among tenants within a high-complexity cloud federation scenario, seems a very challenging task.

A recent piece of work proposes a framework of new metrics able to assess performance and energy efficiency of cloud computing communication systems, processes and protocols (Fiandrino, et al. 2015). However, the authors do not explain how they have obtained their results. This is very difficult for others to replicate and make progress on top of their results. Further work is necessary to standardize the set of metrics that were investigated, and to perform evaluations in operational data centers.

### 4.3 Virtual Networking

The authors of (Panagiotakis, et al. 2015) discuss a potential evolution for the future of mobile multimedia. They predict a networking environment serving a diverse set of pervasive and personalized cloud-based Web applications, where the client-server paradigm will become obsolete. They envision that in the future Web, cloud-based Web applications will be able to communicate, stream and transfer adaptive events and content to their clients, creating a fully collaborative, peer-to-peer and pervasive Web environment. In parallel with these novel requirements, other relevant aspects will also evolve such as the convergence between networking and telecommunications infrastructures, cloud networking, cloud offloading, and the network function virtualization. The new heterogeneous virtualized ecosystem that will be formulated creates new needs and challenges for its management and administration. For this, SDN seems a promising solution (Koumaras, et al. 2015).

A very interesting research direction is the one pointed by (Mastorakis, et al. 2015); this is about the intelligent and efficient management of networking resources on mobile cloud computing (Fernando, Loke and Rahayu 2013). This will be further discussed in the following sub-section.

**4.4 Cooperation, Mobile Cloud Computing, Network Functions Virtualization (NFV), Inter-Cloud Computing Architectures, and Internet of Things (IoT)**

*Cooperation in Cloud Computing*

An obvious method for efficiently using the available cloud resources is to persuade cloud participants to cooperate among themselves. This cooperation can be enforced in several ways: through a common goal (Huerta-Canepa and Lee 2010), using monetary incentives (Charilas, et al. 2011), social incentives (Tanase and Cristea 2011) or reputation incentives (Hwang, Shin and Yoon 2008) (Charilas, et al. 2011). The major problem associated with the common goal method (Huerta-Canepa and Lee 2010) is that it does not work in the absence of a common activity among the potential collaborating entities. In the case of monetary incentive (Charilas, et al. 2011), several issues need to be addressed to identify the most suitable cloud business model to be used (Sen, et al. 2013), and investigate more specific problems such as the credit representation; the security requirements to guarantee a safe monetary transaction; what price to use for each cloud resource; and what type of tariff should be selected (e.g. static, dynamic). Using social incentives such as those suggested in (Tanase and Cristea 2011) also raises some problems such as preventing free riding. The main issues related with reputation mechanisms are the potential lack of fairness and trust associated with the reputation values. This aspect requires further investigation.

*Mobile Cloud Computing and Network Functions Virtualization*

A very significant amount of investigation work has been made in Mobile Cloud Computing (MCC). This can be justified by the exponential increase on handheld mobile devices as well as on the offering of cloud-based services (Wang, Chen, & Wang 2015). The final goal of MCC is to deliver to users a set of mobile services with enhanced QoE. To reach this objective, the mobile operators are deploying an initial strategy that offloads traffic from cellular networks to other available wireless access technologies (e.g. Wifi, WAVE). Other techniques to enhance QoE are service migration and data caching. In this way, and starting with service migration, it can be implemented among federated clouds for offering users a set of services (eventually from distinct cloud providers) with the highest QoE to each user; this offer could be dependent on several requisites namely, the user location, the user profile, and the user terminal characteristics. In addition, the data caching should be deployed to diminish the Round Trip Time (RTT) and its variability (i.e. jitter); consequently, the data should be stored at devices (e.g. MiddleBoxes / Proxies, Access Points, Base Stations, Terminals) very near the user terminals that are expected to consume that data.

To orchestrate all the technologies, strategies and techniques discussed in the last paragraph, making MCC a powerful solution, it is fundamental to program the network and service resources in an intelligent and efficient way. An interesting approach to deliver all this is using Network Functions Virtualization (NFV). The NFV is an emerging network architecture concept that uses virtualization technologies to abstract from the hardware entire classes of network node functions into building blocks that may connect, or chain together, to create intelligent and efficient communication services. As an example, a programmable NFV may consist of one or more virtual machines running, in a coordinated way, different software and processes, on top of standard high-volume servers, switches and storage, or even CC infrastructure, instead of having custom and proprietary hardware appliances for each network function. This new NFV architecture is potentially very flexible; it can deploy virtualized load balancers, firewalls, intrusion detection devices, WAN accelerators, mobile devices power control (Mavromoustakis, et al. 2015), and new MCC business models (Katzis 2015).

The migration of NFV to the cloud environment seems a very challenging task for researchers and engineers due to the myriad of challenges that need to be managed in a harmonized way in order to deliver optimum seamless services to mobile users (Grover and Kheterpal 2015). During, or even better before, the cloudification of NFV services several typical problems associated with mobile networks are urged to be successfully addressed. These problems are related with coverage, interference, congestion and battery autonomy. To solve those problems, various types of resource management techniques should be deployed at mobile clouds such as resource offloading, cloud infrastructure, mobile devices power control (Mavromoustakis, et al. 2015), control theory, data mining, machine learning, radio spectrum management and MCC business models (Katzis 2015).

As a final interesting MCC scenario, (Batalla 2015) elaborates on the particular case related to the delivery of multimedia content to mobile devices originated from media clouds. Since mobile devices are becoming increasingly important receptors of multimedia content, mobile cloud computing is undertaking an important role for delivering audiovisual content from the cloud through the Internet towards the mobile users. On the other hand, high requirements of multimedia content streaming establish the necessity of cross layer mechanisms for avoiding or decreasing the effects of, for example, mobile network congestion or cloud congestion. In this way, one should make use of novel models and algorithms for resource usage prediction that makes possible the optimal distribution of streaming data, and for prediction of the upcoming fluctuations of the network that provide the ability to make the proper decisions in achieving optimized QoS) and QoE for the end users (Kryftis, et al. 2015).

*Inter-Cloud Computing Architectures*

"Storage as a Service" (SaaS) for Internet content delivery, video encoding, and streaming services (e.g. Content Delivery Networks – CDNs) has come to the fore, potentially using a federation of cloud infrastructures. In this context, it is pertinent for providers to hide the different ways in which they operate. One way of performing this transparency is providing a suitable abstraction across the infrastructure heterogeneity. This abstraction can be ensured by a metadata system such as (MetaCDN 2014) (Akamai 2014).

It is also important to be aware of legal issues related to data movement and storage among disparate geographic locations. Notably, the physical locations of both virtual machines and storage arrays have a strong bearing on national laws in respect of security breaches or tampering with data, and in particular where data is moved between different locations (Voorsluys, Broberg and Buyya 2011) (SECCRIT, 2014). There are also important business issues that arise if or when a cloud provider changes owner or closes down, in respect of customer data and applications.

Also recently, research has been carried out in Service-Oriented Architectures (SOAs), especially from a convergence and network point of view (Duan, Yan and Vasilakos 2012). Some relevant aspects of this research will involve several areas, namely network virtualization (Chowdhury and Boutaba 2010) (Jain and Paul 2013) over heterogeneous network infrastructures (e.g. wireless backhaul links, unidirectional optical links) (Tzanakaki, et al. 2013), service discovery technologies (Rambold, et al. 2009), QoS-aware web service composition (Strunk 2010), and network applications based on SDN through a multi-cloud environment (Jain and Paul 2013). SDN has been its main focus in the context of data centers and support of virtualized networks. Consequently, the application of the same approach to wide area networking is still yet to prove its viability. One such application is in supporting lambda path networks, where the elements of the network are not packet switches but wavelength switches (Wei 2014).

A very recent IETF discussion about inter-cloud computing architectures is available in (Aazam, et al. 2015).

*Internet of Things*

The ubiquitous network connectivity, affordable computing power combined with intelligent deployments make the Internet of Things (IoT) very valuable for the current Internet players. The convergence of network wireless access technologies, cloud, and APIs to analyze the data (Big Data) is creating an opportunity for independent software vendors, system integrators, and researchers. Some new solutions are being developed. These solutions are based on new programming models and hardware devices. These can be deployed through very popular languages such as PHP, Python, Java, JavaScript, C#, and Ruby; microcontrollers and low-powered devices such as Arduino, Raspberry Pi, and other embedded devices. The usage scenarios of IoT are diverse and include e-Health, engineering, transportation, and social, to name just a few.

To migrate and operate the IoT devices in the cloud, some obstacles should be overcome; the first challenge is that to realize the true potential of IoT, the data generated by sensors has to be analyzed in real-time; a second challenge is to perform a very useful historical data analysis over structured or even unstructured information previously collected from sensors.

## 5. SECURITY ASPECTS

The topic of security in particular is also discussed in our paper – this has been largely neglected in CC but is beginning to be recognized as a crucial element in the provision of CC services; customers increasingly wish to have assurance that their data and computations will be safe and secure. Trustworthiness is going to be a vital property of CC in the future, especially now that more customers are beginning to place critical services in the Cloud (SECCRIT, 2014). This functional perspective is very pertinent, and needs to be further investigated because the performance of distributed clouds heavily depends on the underlying networks.

A related and important topic is that of resilience – the ability of a system to continue to provide a suitable quality of service even in the face of challenges, when for example security is compromised or a third party event such as a power outage occurs. This is a property that CC systems should strive to provide, especially when supporting critical services (Sterbenz, et al. 2010).

The subsequent discussion is aligned with the aspects identified in Table VIII.

Table VIII. Main Security Aspects of Cloud Systems

| Main Area | Goal |
|---|---|
| Background (Sub-section 5.1) - Generic Cloud Security Aspects | Important security aspects a cloud provider/user should be aware of |
| Background (Sub-section 5.1) - Security Risks Depend Upon the Cloud Service Model | Discussion in how the diverse cloud service models introduce heterogeneous security problems within a cloud system |
| Future Developments (Sub-section 5.2) | Intrusion Detection/Prevention solutions; data privacy; technical and legal issues in CC systems; collusion avoidance mechanisms; secure query over encrypted data |

**5.1 Background**

The relatively new and rapidly adopted model of cloud computing, aggregating in a distributed way so many distinct technologies and solutions, is creating new system vulnerabilities and threats of new and damaging attacks. So, we now also discuss the security aspects of cloud systems.

*Generic Cloud Security Aspects*

Initiating our discussion about security, as a generic (and obvious) but very important topic, one can argue that the security of cloud services should be no worse than that of the network services provided to customers through their local network infrastructures. To achieve this goal, a cloud provider should be conscious of the following aspects:

- The cloud provider needs to apply the most recent security patches in its cloud infrastructure, such as firmware, operating systems and applications. Some problems could occur in the cloud operation due to incompatible patches. In this way, a rollback option should be available to change the infrastructure to the last stable configuration.
- Data isolation must be supported among multiple VMs sharing the resources of the same physical host. Hypervisors also need their security patches to be up to date.
- The cloud paradigm is changing the way the major management networking functions are deployed. These functions running at very specialized equipment located at diverse locations within operators domains, and performing a huge diversity of networking services, such as balancing the load or security, are moving from the operators network core to the cloud (Sherry, et al. 2012). As the cloud infrastructure could be a federation of clouds, then the previous middleboxes should be deployed in a distributed and coordinated way through distinct network domains. This also implies that these middleboxes should be operating with the latest security patches.
- Authentication and trust mechanisms are needed by the user and provider alike. In this scenario, SSO could be a good starting point. The spam e-mail problem can be also mitigated in the cloud (e.g. the spam could be verified and filtered in the VS associated with the hypervisor). Some useful techniques to mitigate spam in clouds could include the Sender Policy Framework (SPF) (Wong and Schlitt 2006) to authenticate the source of each e-mail, and the Apache SpamAssassin Project (SPAMAssassin 2014) to classify, rank and filter any unwanted e-mail.
- To enable communications among the diverse cloud resources/hosts (sometimes from distinct providers) similar to that of a closed local network, the cloud provider's resources/hosts need to be reachable in a secure way through Virtual Private Network (VPN) tunnels. In one initiative, which addresses security and transparency simultaneously, CloudNet makes use of a Virtual Private Cloud (VPC), which brings CC and VPN technology together to give the user a private set of cloud resources (Wood, et al. 2011).

- Cloud services are often made public. Consequently, non-authorized access should be prevented (Patel, et al. 2013) (Modi, et al. 2013). In addition, Distributed Denial of Service (DDoS) attacks carried out by compromised users' machines generate a large amount of bogus traffic. To avoid the negative impact on the system performance of this traffic, the cloud infrastructure can try to identify that traffic and then discard it from the network, redirecting it to a "black hole".
- The Cloud Security Alliance (CSA) is working on the initial identification of top security threats in cloud systems (CSA_a 2014) as well as within mobile computing (CSA_b 2014), and on the establishment of the more convenient actions/strategies to avoid those threats.

*Security Risks Depend Upon the Cloud Service Model*

According to (Fernandes, et al. 2014) (Subashini and Kavitha 2011) it is important that the diverse players using CC should be aware that PaaS, SaaS, and IaaS each have their own security issues. These distinct security aspects are summarized in Table IX, individualized per service model, and discussed in the following text.

Table IX. Main Security Aspects for the Diverse Cloud Service Models

| Service/Topic | Main Aspect | Reference |
| --- | --- | --- |
| SaaS | The SaaS APIs inherit the classical security drawbacks of the Web services | (Fernandes, et al. 2014), (Subashini and Kavitha 2011) |
| PaaS | There is a tradeoff between the level of isolation among tenants and the efficiency level in how the resources are used | (Fernandes, et al. 2014), (Rodero-Merino, et al. 2012) |
| IaaS | Common physical (computing, networking) resources are shared among the customers through virtualized instances | (Fernandes, et al. 2014), (Perez-Botero, et al. 2013), (Vaquero, et al. 2011), (Pearce, et al. 2013) |

The SaaS services are very similar to Web services over HTTP. In this way, the former inherits the classical security drawbacks of the latter, as follows:
- The SaaS interface can be maliciously hacked through application loopholes (i.e. vulnerability in the system that enables an attacker to compromise that system) (Subashini and Kavitha 2011).
- The attacker can inject masked code into a SaaS system that can break isolation barriers (Subashini and Kavitha 2011).
- The lack of data integrity in the messages such that it can be changed during their transmission through the network in favor of a particular malicious intent of a man-in-the-middle attacker (Fernandes, et al. 2014).

The PaaS systems are based on platforms such as, .NET and Java. The resources offered by these platforms are shared among multiple customers (i.e. multitenancy aspect). Consequently, a proper isolation mechanism must ensure that one tenant cannot access to components of other tenants. For this, there is a clear tradeoff between resource consumption and the isolation level to be offered. Further discussion on this is in (Rodero-Merino, et al. 2012).

Common physical (computing, networking) resources are shared among the customers through virtualized instances, offering IaaS solutions. (Vaquero, et al. 2011) discussed security from the networking, virtualization and physical sides of cloud IaaS networks. There are also management consoles, such as XenCenter for Xen VMs, which can be remotely accessed via the Web. Consequently, these management consoles are also vulnerable to a VM-to-VMM attack that consists in gaining access to the underlying VMM (e.g. VmwarePlayer, VirtualBox) through a legitimately running VM managed by that VMM. This attack is normally designated by VM escape (Grobauer, et al. 2011). If this attack is successful, the attacker can monitor other VMs, including shared resources and CPU utilization, and shutting down VMs. In respect to the networking aspect, the VMMs typically offer various basic types of networking to child VMs (Pearce, et al. 2013): bridging virtual NICs to physical adapters (appears to be directly connected to the physical network), Network Address Translation (NAT) routing (sharing the IP address of the host), and internal and isolated networking (private network shared with the host). On public IaaS clouds, it is desirable to treat VMs as if they are standard physical servers, thereby bridging VMs networking seeming as the better solution. A bridged adapter can capture traffic on the physical network, without any control from the physical host. This can be an issue in case of promiscuous mode where VMs can analyze all traffic including that not addressed to them (Pearce, et al. 2013). To aggravate the scenario, VMMs are known not to yet be bug-free and, from time to time, a vulnerability comes along, as surveyed by (Perez-Botero, et al. 2013), who presented lists of vulnerabilities for Xen and KVM.

**5.2 Future Developments**

There are also some available surveys concerning security issues in CC (Patel, et al. 2013) (Modi, et al. 2013) (Subashini and Kavitha 2011), namely the ones that can impair integrity, availability, and confidentiality. Using only firewall devices will not help solve these problems. Consequently, (Patel, et al. 2013) (Modi, et al. 2013) examine proposals that incorporate the joint use of IDS (Intrusion Detection Systems) and IPS (Intrusion Prevention Systems). Finally, (Samanthula, et al. 2015) (Fernandes, et al. 2014) discuss threats coming from the diversity of the SaaS, PaaS and IaaS approaches. They also discuss some solutions to target the security challenges in clouds. On one hand, the proposals based on signature detection offer the advantage of minimal response time and human intervention but have the disadvantage of not being able to detect previously unknown ('zero day') attacks. On the other hand, anomaly detection proposals have opposite functional characteristics in comparison with signature-based ones. Hybrid cloud IDPS schemes should be investigated for use in future systems.

Future cloud systems should be able to detect and prioritize simultaneous attacks in terms of their negative impact on the system performance. Then, these systems need to put into action prioritized corrective measures to limit the destructiveness of the more dangerous attacks. In addition, the security solutions should scale or adjust network node numbers, the node heterogeneity (e.g. a federated cloud system), and traffic load, to offer a satisfactory service. It is also worth noting that there is a trade-off between performance and the level of security adopted. Clearly, higher security levels will necessitate more checking, and consequently there will be fewer resources for regular customer use. It is therefore advisable to apply the minimally appropriate set of policies by means of self-managing and self-learning.

Cloud users would also need to feel confident that their data privacy is guaranteed when they upload the data to the cloud. To address this security requirement, as suggested in (Satyanarayanan, et al. 2009), would require trust establishment methods.

A significant piece of research is currently being carried out in the European FP7 project SECCRIT (Secure Cloud Computing for Critical Infrastructure IT), which addresses technical and legal issues in the context of cloud security. This (SECCRIT, 2014) (Bless, et al. 2013) "is a multidisciplinary research project with the mission to analyze and evaluate cloud computing technologies with respect to security risks in sensitive environments, and to develop methodologies, technologies, and best practices for creating secure, trustworthy, and high assurance cloud computing for critical infrastructure IT." Also, the project is investigating relevant European legal frameworks with the aim of establishing guidelines for using cloud services in the critical infrastructure sector. Otherwise, the use of cloud in this sector, where stringent regulatory and legal requirements exist, will continue to be severely limited. Furthermore, clear guidelines are needed on how to deal with liability issues following any service failures.

Very recently a new cloud service model is winning a considerable importance, the Data as a Service (DaaS), which we discuss in the following subsection.

*Security Risks of an Emerging Cloud Service Model: Data as a Service*

A very recent piece of work (Samanthula, et al. 2015) complements previous work (Fernandes, et al. 2014) (Subashini and Kavitha 2011), discussing the security risks involved with an emerging cloud service model: Data as a Service (DaaS). The typical usage scenario of this model is the one where the user data is outsourced to the cloud (e.g. Dropbox). However, the data owners lose control over their data because the cloud provider becomes a third party service provider. An initial solution to ensure the data privacy is to encrypt it before exporting it to the cloud. A legacy solution to this issue is based on symmetric key encryption but it is not secure when a revoked user rejoins the system. In this way, (Samanthula, et al. 2015) proposes a homomorphic encryption and proxy re-encryption scheme that prevents leakage of data privacy when a revoked user rejoins the system. This solution also prevents the collusion between a revoked user and the cloud provider. It also supports secure query processing over the encrypted data already stored in a federation of clouds. Further information on this is available in (Samanthula, et al. 2015).

## 6. OPEN ISSUES

We now highlight some unresolved issues and point out future networking research directions in the area of CC (Table X).

The first key issue is the dynamic management of cloud resources in resource-constrained scenarios (Bodik, et al. 2012) (Raiciu, et al. 2011) (Detal, et al. 2013) or federated environments with service migration (Popa, et al. 2011) (MetaCDN 2014) (Akamai 2014). This resource management needs to be balanced against other aspects, notably fault tolerance (Bodik, et al. 2012), energy consumption (Voorsluys, Broberg and Buyya 2011) (Sakamoto, et al. 2012), network utilization (Raiciu, et al. 2011), load balancing (Detal, et al. 2013), data congestion (Popa, et al. 2011), and data availability (MetaCDN 2014) (Akamai 2014). As an example, SDN may be used to limit the packet flow rate and to forward intelligently the data packets using convenient management policies, respectively, to mitigate congestion and optimize the data availability. In addition, another very interesting challenge needs to be addressed, namely the efficient delivery of diverse services, such as computation, storage, virtualization, applications, and networks (Buyya 2014).

SDN can be also useful for enhancing the available security in cloud environments, e.g. data centers, by deploying new features such as IDPS (Patel, et al. 2013) (Modi, et al. 2013). It is also important to combine research on legal aspects alongside those of security and resilience if CC and services are to be successfully deployed in critical infrastructure IT (SECCRIT, 2014) (Sterbenz, et al. 2010). There are a few open issues that need to be addressed for providing a secure CC environment (Ali, Khan, & Vasilakos 2015), such as:

- Harmonizing different security solutions within the cloud systems to offer the desired security level.
- Addressing multi tenancy security issues, namely to ensure the privacy during computations within virtualized, shared and distributed processing environments.

- Security against insider threats; these insider attacks can be avoided to an extent by having definitive criteria for judging between normal and malicious (or compromised) user behavior.
- Finding solutions that create a proper balance between the security requirements and cloud performance.

Table X. Summary of Open Networking-Based Issues to Deploy Cloud Computing in Future Networks

| Open Issue | Reference |
| --- | --- |
| Dynamic management of cloud resources | (Bodik, et al. 2012) (Raiciu, et al. 2011) (Detal, et al. 2013) |
| Cloud federation environments | (Popa, et al. 2011) (MetaCDN 2014) (Akamai 2014) |
| Fault tolerance | (Bodik, et al. 2012) |
| Energy consumption | (Voorsluys, Broberg and Buyya 2011) (Sakamoto, et al. 2012) |
| Network utilization | (Raiciu, et al. 2011) |
| Load balancing | (Detal, et al. 2013) |
| Data congestion | (Popa, et al. 2011) |
| Data availability | (MetaCDN 2014) (Akamai 2014) |
| Intrusion detection and prevention systems | (Patel, et al. 2013) (Modi, et al. 2013) |
| Legal aspects alongside security and resilience | (SECCRIT, 2014) |
| Harmonize a large number of diverse security solutions; address multi tenancy security issues; mitigate insider attacks; ensure the right balance between security efficiency and cloud performance | (Ali, Khan, & Vasilakos 2015) |
| Outsourcing of middleboxes (e.g. NATs, firewalls, load balancers) to the cloud; optimizing mobile networks through the management of flows | (Sherry, et al. 2012) (Silva, et al. 2013) |
| Network hypervisors (i.e. hypervisors coupled with virtual switches controlled by SDN) | (Vmware NSX 2014) |
| CC and Internet of Things | (Comer, 2014) |
| The new-business potential of clouds | (Sen, et al. 2013) (Berman, et al. 2012) (Sharkh, et al. 2013) |

As suggested in (Sherry, et al. 2012), it will be necessary to investigate the outsourcing of middleboxes (e.g. NATs, firewalls, load balancers) to the cloud. This outsourcing is justified by the fact the current middleboxes being deployed within the networks of customers impose a considerable cost, management complexity and network overhead. In addition, network hypervisors (i.e. hypervisors coupled with vSwitches controlled by SDN) can bring to future networks the benefits of machine virtualization in terms of flexibility, scale, performance, and assurance, by creating a virtualized network infrastructure (Vmware NSX 2014). This is provisioned as an overlay solution that offers to the application level a full set of reliable networking services with complete independence of both the underlying network layers (router/switch hardware, physical network topology) and operator domains. (Silva, et al. 2013) also proposed, at the network edge, a solution that controls the admission of mobile flows in a resource-constrained scenario. Additionally, the accepted flows are managed according to their Classes of Service. The output of this last work could be particularly interesting to be adopted in MCC scenarios.

A new networking paradigm is showing up, namely intelligent embedded systems sensing of local information and reporting it to the Internet for further analysis. Researchers are using the term Internet of Things (IoT) to designate this emerging area. This model should be very relevant everywhere, e.g. in smart cities, houses, office buildings, vehicles, shopping malls, and industrial applications (Comer, 2014). The exponential proliferation of these small devices, each one requiring an IP address for communication with specialized CC systems, should at long last help accelerate the adoption of IPv6.

Companies across the globe clearly also see the cloud's new-business potential (Sen, et al. 2013) for promoting sustainable competitive advantage against their market competitors (Berman, et al. 2012).

In summary, the providers, developers, and end-users of CC must consider several issues in order to take best advantage of CC; these including security, privacy, trust, and resilience; interoperability among distinct CC infrastructures; availability, fault-tolerance, and disaster recovery; and resource management. Another very important CC challenge to be addressed is the 'green' aspect of power efficiency in cloud systems (Sharkh, et al. 2013). If these diverse cloud challenges and risks are correctly addressed by industry and academia, possibly working in tandem, the long-term success of CC will hopefully be guaranteed (Voorsluys, Broberg and Buyya 2011).

## 7. CONCLUSION

Despite the many advantages offered by CC, there are also networking concerns that hamper its fast adoption. This article has reviewed and analyzed the networking-related issues that arise due to resource outsourcing, the virtualized, shared, and public nature of CC, the emerging challenges from security breaches, and the increasing need to provide a resilient CC infrastructure and services.

The major goal of this article was to examine comprehensively the role of networking in CC, and the issues arising. We looked at the origins of CC and discussed the various developments that brought it to the present day. Foundation technologies and architectural models were discussed, as well as some of the more relevant CC offerings. The most pertinent network aspects were presented and discussed in detail, focusing on the crucial support that the networking infrastructure provides for CC. This discussion also presented and examined relevant contributions from industry, academia and standardization arenas. Finally, the article also highlighted relevant CC areas requiring further research.

## ACKNOWLEDGMENT

The research presented in this article was partly funded by the European Union Seventh Framework Programme (FP7/2007-13), grant agreement no. 312758: the SECCRIT project.

## REFERENCES

Agarwal, Sharad, John Dunagan, Navendu Jain, Stefan Saroiu, Alec Wolman, and Harbinder Bhogan. 2010. Volley: Automated Data Placement for Geo-distributed Cloud Services. Proceedings of the 7th USENIX Conference on Networked Systems Design and Implementation. Berkeley, CA, USA: USENIX Association, 16 pages.

Alamri, Atif, Wasai Shadab Ansari, Mohammad Mehedi Hassan, M. Shamim Hossain, Abdulhameed Alelaiwi, and M. Anwar Hossain. 2013. A Survey on Sensor-Cloud: Architecture, Applications, and Approaches. International Journal of Distributed Sensor Networks, 18 pages.

Ali, M., Khan, S., Vasilakos, A.. 2015. Security in cloud computing: Opportunities and challenges, Information Sciences, 305(1), 357-383.

Amamou, Ahmed, Haddadou, Kamel, Pujolle, Guy. 2014. A TRILL-based multitenant data center network, Computer. Networks, 68, 35-53.

Amazon. Elastic Compute Cloud. 2013. https://aws.amazon.com/pt/ec2/ (retrieved 02/03/2014).

Amiri, Khalil, David Petrou, Gregory R. Ganger, and Garth A. Gibson. 2000. Dynamic Function Placement for Data-intensive Cluster Computing. Proceedings of the Annual Conference on USENIX Annual Technical Conference. Berkeley, CA, USA: USENIX Association, 16 pages.

ANSI/INCITS. Fibre Channel Backbone-5 Rev 2.0. ANSI. 04/06/2009. http://www.t11.org/ftp/t11/pub/fc/bb-5/09-056v5.pdf (retrieved 03/03/2014).

Armbrust, Michael et al. 2009. Above the Clouds: A Berkeley View of Cloud Computing. Technical Report No. UCB/EECS-2009-28, 23 pages, 2009.

Armbrust, Michael et al. 2010. A View of Cloud Computing. Commununications of the ACM (ACM) 53, n.º 4, 50-58.

AT&T. AT&T Cloud Architect. 2012. http://cloudarchitect.att.com/Home/ (retrieved 02/03/2014).

Aazam, M., Huh, E.-N., Kim, S.. 2015. Inter-Cloud Computing Architecture, IETF Informational document, draft-aazam-cdni-inter-cloud-architecture-02 (Expires in 17/09/2015), 22 pages.

Azodolmolky, S., P. Wieder, and R. Yahyapour. 2013. Cloud computing networking: challenges and opportunities for innovations. IEEE Communications Magazine 51, n.º 7 (July 2013), 54-62.
Badger, Lee, Tim Grance, Robert Patt-Comer, and Jeff Voas. 2012. Cloud Computing Synopsis and Recommendations, NIST Special Publication 800-146. NIST. May of 2012. http://csrc.nist.gov/publications/nistpubs/800-146/sp800-146.pdf (retrieved 03/03/2014).
Ballani, Hitesh, Paolo Costa, Thomas Karagiannis, and Ant Rowstron. 2011. Towards Predictable Datacenter Networks. Proceedings of the ACM SIGCOMM 2011 Conference. New York, NY, USA: ACM, 2011, 242-253.
Banikazemi, M., Olshefski, D., Shaikh, A., Tracey, J., Wang, G.. 2013. Meridian: an SDN platform for cloud network services, IEEE Communications Magazine, 51( 2), 120–127.
Bansal, N., R. Bhagwan, N. Jain, Yoonho Park, D. Turaga, and C. Venkatramani. 2008. Towards Optimal Resource Allocation in Partial-Fault Tolerant Applications. INFOCOM 2008. The 27th Conference on IEEE Computer Communications. 2008. 10 pages.
Batalla, J. M. (2015). Adaptation of Cloud Resources and Media Streaming in Mobile Cloud Networks for Media Delivery. In G. Mastorakis, C. Mavromoustakis, & E. Pallis (Eds.) Resource Management of Mobile Cloud Computing Networks and Environments (pp. 175-202). Hershey, PA: Information Science Reference. doi:10.4018/978-1-4666-8225-2.ch007
Beloglazov, Anton, Rajkumar Buyya, Young Choon Lee, and Albert Y. Zomaya. 2011. A Taxonomy and Survey of Energy-Efficient Data Centers and Cloud Computing Systems. Advances in Computers 82 (2011), 47-111.
Benson, T., Akella, A., Shaikh, A., Sahu, S.. 2011. Cloudnaas: a cloud networking platform for enterprise applications. In Proceedings of the 2nd ACM Symposium on Cloud Computing, ser. SOCC '11, 8:1–8:13.
Birke, R., L.Y. Chen, and E. Smirni. 2012. Data Centers in the Cloud: A Large Scale Performance Study. IEEE 5th International Conference on Cloud Computing (CLOUD). 2012, 336-343.
Bless, Roland, David Hutchison, Marcus Schöller, Paul Smith, and Markus Tauber. SECCRIT: Secure Cloud Computing for High Assurance Services. *ERCIM News* 95, 2013.
Bodik, Peter, Ishai Menache, Mosharaf Chowdhury, Pradeepkumar Mani, David A. Maltz, and Ion Stoica. 2012. Surviving Failures in Bandwidth-constrained Datacenters. Proceedings of the ACM SIGCOMM 2012 Conference on Applications, Technologies, Architectures, and Protocols for Computer Communication. New York, NY, USA: ACM, 2012, 431-442.
Boutaba, Raouf, Limam, Noura, Secci, Stefano, Taleb, Tarik. 2014. Cloud networking and communications. Computer Networks, 68, 1-4.
Briscoe, Bob, and M. Sridharan. IETF - Internet Draft. IETF. 14/02/2014. http://tools.ietf.org/html/draft-briscoe-conex-data-centre-02 (retrieved 03/03/2014).
BT. Cloud Compute. 2014. http://www.globalservices.bt.com/uk/en/products/cloud_compute (retrieved 02/03/2014).
Buyya, Rajkumar, Chee Shin Yeo, Srikumar Venugopal, James Broberg, and Ivona Brandic. 2009. Cloud Computing and Emerging IT Platforms: Vision, Hype, and Reality for Delivering Computing As the 5th Utility. Future Generation Computer Systems (Elsevier Science Publishers B. V.) 25, n.º 6 (june 2009), 599-616.
Buyya, Rajkumar. 2014. Introduction to the IEEE Transactions on Cloud Computing, IEEE Transactions on Cloud Computing, vol. 1, no. 1, 3-9.
CCGa. Cloud Computing Glossary (1). 2014. http://cloudtimes.org/glossary/ (retrieved 02/03/2014).
CCGb. Cloud Computing Glossary (2). 2014. http://cloudglossary.com/ (retrieved 02/03/2014).
Chowdhury, N.M., Mosharaf Kabir, and Raouf Boutaba. 2010. A Survey of Network Virtualization. Computer Networks (Elsevier North-Holland, Inc.) 54, n.º 5 (april 2010), 862-876.
Charilas, D., Stavroula Vassaki, Athanasios Panagopoulos, Philipp Constantinou. 2011. Cooperation Incentives in 4G Networks. (Eds.) Y. Zhang, M. Guizani. Game Theory for Wireless Communications and Networking, CRC Press, Chap. 13, 295-314.
Comer, Douglas. 2014. The ZigBee IP Protocol Stack. The Internet Protocol Journal, vol. 17, no. 2. December of 2014. http://ipj.dreamhosters.com/wp-content/uploads/2014/12/ipj17.2.pdf, 19-38.
Corradi, A., Fanelli, M., Foschini, L.. 2014. VM consolidation: A real case based on openstack cloud, Future Generation Computer Systems, 32( 0), 118–127.
DARPA_a. Proceed Darpa Project. DARPA. 2013. http://www.darpa.mil/Our_Work/I2O/Programs/PROgramming_Computation_on_EncryptEd_Data_(PROCEED).aspx (retrieved 02/03/2014).
DARPA_b. DARPA. Mission-oriented Resilient Clouds (MRC). 2013. http://www.darpa.mil/Our_Work/I2O/Programs/Mission-oriented_Resilient_Clouds_(MRC).aspx (retrieved 02/03/2014).
Detal, Gregory, Christoph Paasch, Simon Van Der Linden, Pascal Merindol, Gildas Avoine, and Olivier Bonaventure. 2013. Revisiting Flow-based Load Balancing: Stateless Path Selection in Data Center Networks. Computer Networks (Elsevier North-Holland, Inc.) 57, n.º 5 (april 2013), 1204-1216.
Diebold, Francis. On the Origin(s) and Development of the Term 'Big Data'. PIER Working Paper No. 12-037. 21/09/2012. http://papers.ssrn.com/sol3/papers.cfm?abstract_id=2152421 (retrieved 02/03/2014).
Dinh, Hoang T., Chonho Lee, Dusit Niyato, and Ping Wang. 2013. A survey of mobile cloud computing: architecture, applications, and approaches. Wireless Communications and Mobile Computing 13, n.º 18 (2013), 1587-1611.
Dropbox. Dropbox. 2014. https://www.dropbox.com/ (retrieved 02/03/2014).

DT. Cloud. 2014. http://www.telekom.com/innovation/80328 (retrieved 02/03/2014).
Duan, Qiang, Yuhong Yan, and A.V. Vasilakos. 2012. A Survey on Service-Oriented Network Virtualization Toward Convergence of Networking and Cloud Computing. IEEE Transactions on Network and Service Management 9, n.º 4 (December 2012), 373-392.
Duffield, N. G., Pawan Goyal, Albert Greenberg, Partho Mishra, K. K. Ramakrishnan, and Jacobus E. van der Merive. 1999. A Flexible Model for Resource Management in Virtual Private Networks. Proceedings of the Conference on Applications, Technologies, Architectures, and Protocols for Computer Communication. New York, NY, USA: ACM, 1999, 95-108.
ETSI. American and European Standards Organizations agree to collaborate on aligning standards to facilitate trade between EU and US. ETSI. 14/02/2013. http://www.etsi.org/news-events/news/649-2013-02-ansi-eso-collaboration-at-jpg?highlight=YTozOntpOjA7czo1OiJjbG91ZCI7aToxO3M6OToiY29tcHV0aW5nIjtpOjI7czoxNToiY2xvdWQgY29tcHV0aW5nIjt9 (retrieved 02/03/2014).
Eucalyptus. Eucalyptus. Eucalyptus. 2014. https://www.eucalyptus.com/why-eucalyptus (retrieved 03/03/2014).
Fernandes, D., et al.. 2014. Security issues in cloud environments: a survey. Int. J. Inform. Sec. 13 (2), 113–170.
Fernando, Niroshinie, Seng W. Loke, and Wenny Rahayu. 2013. Mobile Cloud Computing: A Survey. Future Generation Computer Systems (Elsevier Science Publishers B. V.) 29, n.º 1 (#jan# 2013): 84-106.
Fiandrino, C., Kliazovich, D., Bouvry, P., Zomaya, A. 2015. Performance and Energy Efficiency Metrics for Communication Systems of Cloud Computing Data Centers, IEEE Transactions on Cloud Computing, PP (99), 14 pages.
Gartner. Gartner Says Worldwide Public Cloud Services Market to Total $131 Billion. 28/02/2013. http://www.gartner.com/newsroom/id/2352816 (retrieved 02/03/2014).
Google_a. Google App Engine: Platform as a Service. 07/02/2014. https://developers.google.com/appengine/ (retrieved 02/03/2014).
Google_b. Google Drive. 2014. http://learn.googleapps.com/drive (retrieved 02/03/2014).
GridGain. GridGain In-Memory Computing. 2014. http://www.gridgain.com/ (retrieved 02/03/2014).
Griebel L., Prokosch H.-U., Köpcke F., Toddenroth D., Christoph J., Leb I., Engel, I., Sedlmayr, M. (2015). A scoping review of cloud computing in healthcare. BMC Med. Inform. Decis. Mak. 15 (17), 10.1186/s12911-015-0145-7, 16 pages.
Grobauer, B., Walloschek, T., Stocker, E.. 2011. Understanding Cloud Computing Vulnerabilities. IEEE Secur. Privacy 9(2), 50-57.
Grover, J., & Kheterpal, G. (2015). Mobile Cloud Computing: An Introduction. In G. Mastorakis, C. Mavromoustakis, & E. Pallis (Eds.) Resource Management of Mobile Cloud Computing Networks and Environments (pp. 1-23). Hershey, PA: Information Science Reference. doi:10.4018/978-1-4666-8225-2.ch001
Gupta, A., et al.. 2014. SDX: a software defined internet exchange. SIGCOMM Comput. Commun. Rev. 44, 4 (August 2014), 551-562.
Hu, Yan, Ming Zhu, Yong Xia, Kai Chen, and Yanlin Luo. 2012. GARDEN: Generic Addressing and Routing for Data Center Networks. IEEE 5th International Conference on Cloud Computing (CLOUD). 2012, 107-114.
Huerta-Canepa, Gonzalo, and Dongman Lee. 2010. A Virtual Cloud Computing Provider for Mobile Devices. Proceedings of the 1st ACM Workshop on Mobile Cloud Computing Services: Social Networks and Beyond. New York, NY, USA: ACM, 2010, 6:1--6:5.
Huston, Geoff. A Retrospective: Twenty-Five Years Ago. The Internet Protocol Journal, 15 (1). March of 2012. http://www.cisco.com/web/about/ac123/ac147/archived_issues/ipj_15-1/151_25-years-ago.html, 24-35.
Hwang, Junseok, Andrei Shin, and Hyenyoung Yoon. 2008. Dynamic Reputation-based Incentive Mechanism Considering Heterogeneous Networks. Proceedings of the 3Nd ACM Workshop on Performance Monitoring and Measurement of Heterogeneous Wireless and Wired Networks. New York, NY, USA: ACM, 2008, 137-144.
IEEE. IEEE Standard for Local and metropolitan area networks-- Virtual Bridged Local Area Networks Amendment 13: Congestion Notification. IEEE Std 802.1Qau-2010 (Amendment to IEEE Std 802.1Q-2005), April 2010: c1-119.
IEEE. IEEE Standard for Local and metropolitan area networks--Media Access Control (MAC) Bridges and Virtual Bridged Local Area Networks--Amendment 17: Priority-based Flow Control. IEEE Std 802.1Qbb-2011 (Amendment to IEEE Std 802.1Q-2011 as amended by IEEE Std 802.1Qbe-2011 and IEEE Std 802.1Qbc-2011), Sept 2011, 1-40.
Jain, R., and S. Paul. 2013. Network virtualization and software defined networking for cloud computing: a survey. IEEE Communications Magazine 51, n.º 11 (November 2013), 24-31.
Jamshidi, Pooyan, Ahmad, Aakash, and Pahl, Claus. 2013. Cloud migration research: A systematic review. IEEE Trans. Cloud Comput., vol. 1, no. 2, 142–157.
Jin, Hai, Shadi Ibrahim, Li Qi, Haijun Cao, Song Wu, and Xuanhua Shi. The MapReduce Programming Model and Implementations. In Cloud Computing, 373-390. John Wiley & Sons, Inc.,

2011.
Katzis, K. (2015). Mobile Cloud Resource Management. In G. Mastorakis, C. Mavromoustakis, & E. Pallis (Eds.) Resource Management of Mobile Cloud Computing Networks and Environments (pp. 69-96). Hershey, PA: Information Science Reference. doi:10.4018/978-1-4666-8225-2.ch004

Keahey, K., I. Foster, T. Freeman, and X. Zhang. 2005. Virtual Workspaces: Achieving Quality of Service and Quality of Life in the Grid. Scientific Programming (IOS Press) 13, n.º 4 (#oct# 2005), 265-275.

Kim, Hyunjoo, and Manish Parashar. 2011. CometCloud: An Autonomic Cloud Engine. In Cloud Computing, 275-297. John Wiley & Sons, Inc., 2011.

Koumaras, H., Damaskos, C., Diakoumakos, G., Kourtis, M., Xilouris, G., Gardikis, G., Koumaras, V., & Siakoulis, T. (2015). Virtualization Evolution: From IT Infrastructure Abstraction of Cloud Computing to Virtualization of Network Functions. In G. Mastorakis, C. Mavromoustakis, & E. Pallis (Eds.) Resource Management of Mobile Cloud Computing Networks and Environments (pp. 279-306). Hershey, PA: Information Science Reference. doi:10.4018/978-1-4666-8225-2.ch010

Kryftis, Y., Mastorakis, G., Mavromoustakis, C. X., Batalla, J. M., Bourdena, A., & Pallis, E. (2015). A Resource Prediction Engine for Efficient Multimedia Services Provision. In G. Mastorakis, C. Mavromoustakis, & E. Pallis (Eds.) Resource Management of Mobile Cloud Computing Networks and Environments (pp. 361-380). Hershey, PA: Information Science Reference. doi:10.4018/978-1-4666-8225-2.ch012

Li, Feng, Beng Chin Ooi, M. Tamer Ozsu, and Sai Wu. 2014. Distributed Data Management Using MapReduce. ACM Computing Surveys (ACM) 46, n.º 3 (#jan# 2014), 31:1--31:42.

Mastorakis, G., Mavromoustakis, C. X., & Pallis, E. (2015). Resource Management of Mobile Cloud Computing Networks and Environments (pp. 1-432). Hershey, PA: IGI Global. doi:10.4018/978-1-4666-8225-2

Mavromoustakis, C. X., Mastorakis, G., Bourdena, A., Pallis, E., Stratakis, D., Perakakis, E., Kopanakis, I., Papadakis, S., Zaharis, Z. D., Skeberis, C., & Xenos, T. D. (2015). A Social-Oriented Mobile Cloud Scheme for Optimal Energy Conservation. In G. Mastorakis, C. Mavromoustakis, & E. Pallis (Eds.) Resource Management of Mobile Cloud Computing Networks and Environments (pp. 97-121). Hershey, PA: Information Science Reference. doi:10.4018/978-1-4666-8225-2.ch005

McKeown, Nick et al.. 2008. OpenFlow: Enabling Innovation in Campus Networks. SIGCOMM Computer Communication Review (ACM) 38, n.º 2 (#mar# 2008), 69-74.

Medina, Violeta, and Juan Manuel Garcia. 2014. A Survey of Migration Mechanisms of Virtual Machines. ACM Computing Surveys (ACM) 46, n.º 3 (#jan# 2014), 30:1-30:33.

Mei, Lijun, W.K. Chan, and T. H. Tse. 2008. A Tale of Clouds: Paradigm Comparisons and Some Thoughts on Research Issues. Asia-Pacific Services Computing Conference, 2008. APSCC '08. IEEE. 2008, 464-469.

Mell, Peter, and Timothy Grance. The NIST Definition of Cloud Computing. Tech. rep., National Institute of Standards and Technology (NIST), Gaithersburg, MD, 2011.

Metri, G., S. Srinivasaraghavan, Weisong Shi, and M. Brockmeyer. 2012. Experimental Analysis of Application Specific Energy Efficiency of Data Centers with Heterogeneous Servers. IEEE 5th International Conference on Cloud Computing (CLOUD). 2012, 786-793.

Microsoft_a. Windows Azure. 2014. https://www.windowsazure.com/en-us/ (retrieved 02/03/2014).

Microsoft_b. OneDrive. 2014. https://onedrive.live.com/about/en-us/ (retrieved 02/03/2014).

Modi, Chirag, Dhiren Patel, Bhavesh Borisaniya, Hiren Patel, Avi Patel, and Muttukrishnan Rajarajan. 2013. A survey of intrusion detection techniques in Cloud. Journal of Network and Computer Applications 36, n.º 1 (#jun# 2013), 42-57.

Mohapatra, P., et al.. 2013. Fast Connectivity Restoration Using BGP Add-path. IETF - Internet Draft. IETF. https://tools.ietf.org/html/draft-pmohapat-idr-fast-conn-restore-03 (retrieved 26/06/2015).

Moura, Jose, and Serrão, Carlos. (2015). Security and Privacy Issues of Big Data. In N. Zaman, M. Seliaman, M. Hassan, & F. Marquez (Eds.) Handbook of Research on Trends and Future Directions in Big Data and Web Intelligence. Hershey, PA: Information Science Reference. doi:10.4018/978-1-4666-8505-5.ch002, 20-52.

MT. SAAS Application Hosting. 2014. http://www.macquarietelecom.com/solutions/pure-saas-application-hosting/ (retrieved 02/03/2014).

Murphy, MichaelA., Linton Abraham, Michael Fenn, and Sebastien Goasguen. 2010. Autonomic Clouds on the Grid. Journal of Grid Computing (Springer Netherlands) 8, n.º 1 (2010), 1-18.

ND. Cloud Solutions. 2014. http://www.nttdata.com/global/en/services/cloud/index.html (retrieved 02/03/2014).

Nunes, B., M. Mendonca, X. Nguyen, K. Obraczka, and T. Turletti. 2014. A Survey of Software-Defined Networking: Past, Present, and Future of Programmable Networks. IEEE Communications Surveys Tutorials PP, n.º 99 (2014), 1-18.

Obama. Big Data Initiative: Announces $200 Million in new R&D Investments. 29/03/2012. http://www.whitehouse.gov/sites/default/files/microsites/ostp/big_data_press_release_final_2.pdf (retrieved 02/03/2014).

Oracle. Cloud Computing. May of 2010. http://www.oracle.com/us/technologies/cloud/oracle-cloud-computing-wp-076373.pdf (retrieved 02/03/2014).

Panagiotakis, S., Vakintis, I., Andrioti, H., Stamoulias, A., Kapetanakis, K., & Malamos, A. (2015). Towards Ubiquitous and Adaptive Web-Based Multimedia Communications via the Cloud. In G. Mastorakis, C. Mavromoustakis, & E. Pallis (Eds.) Resource Management of Mobile Cloud Computing Networks and Environments (pp. 307-360). Hershey, PA: Information Science

Reference. doi:10.4018/978-1-4666-8225-2.ch011

Patel, Ahmed, Mona Taghavi, Kaveh Bakhtiyari, and Joaquim Celestino Junior. 2013. Review: An Intrusion Detection and Prevention System in Cloud Computing: A Systematic Review. Journal of Network and Computer Applications (Academic Press Ltd.) 36, n.º 1 (#jan# 2013), 25-41.

Pearce, M., Zeadally, S., Hunt, R.. 2013. Virtualization: Issues, Security Threats, and Solutions. ACM Comput. Surv. 45(2), 17:1-17:39

Perez-Botero, D., Szefer, J., Lee, R.B. 2013. Characterizing Hypervisor Vulnerabilities in Cloud Computing Servers. In: Proc. of the Int. Workshop on Security in Cloud Computing (SCC), pp. 3-10.

Popa, Lucian, Arvind Krishnamurthy, Sylvia Ratnasamy, and Ion Stoica. 2012. FairCloud: sharing the network in cloud computing. SIGCOMM 2012. ACM, 2012, 12 pages.

Press, Gil. A Very Short History Of Big Data. Forbes. 21/12/2013. http://www.forbes.com/sites/gilpress/2013/05/09/a-very-short-history-of-big-data/ (retrieved 02/03/2014).

ProgrammableWeb. Programmable Web. 2014. http://www.programmableweb.com/ (retrieved 03/03/2014).

PT. Cloud Solutions. 2014. https://cloud.ptempresas.pt/Pages/Catalog/ServiceDetail.aspx?s=06IG3nFf0pSkKNHn-KBVCw&language=en-US (retrieved 02/03/2014).

Raiciu, Costin, Sebastien Barre, Christopher Pluntke, Adam Greenhalgh, Damon Wischik, and Mark Handley. 2011. Improving Datacenter Performance and Robustness with Multipath TCP. Proceedings of the ACM SIGCOMM 2011 Conference. New York, NY, USA: ACM, 2011, 266-277.

Rambold, Michael, Holger Kasinger, Florian Lautenbacher, and Bernhard Bauer. 2009. Towards Autonomic Service Discovery A Survey and Comparison. Proceedings of the 2009 IEEE International Conference on Services Computing. Washington, DC, USA: IEEE Computer Society, 2009, 192-201.

Retana, A., White, R. The Internet Protocol Journal, vol. 16, no.2., June of 2013.http://www.cisco.com/web/about/ac123/ac147/archived_issues/ipj_16-2/ipj_16-2.pdf, 23-29.

Ricci, Robert, Chris Alfeld, and Jay Lepreau. 2003. A Solver for the Network Testbed Mapping Problem. SIGCOMM Computer Communication Review (ACM) 33, n.º 2 (#apr# 2003), 65-81.

Rimal, Bhaskar Prasad, Eunmi Choi, and Ian Lumb. 2009. A Taxonomy and Survey of Cloud Computing Systems. Proceedings of the 2009 Fifth International Joint Conference on INC, IMS and IDC. Washington, DC, USA: IEEE Computer Society, 2009, 44-51.

Rodero-Merino, L., Vaquero, L.M., Caron, E., Desprez, F., Muresan, A.. 2012. Building Safe PaaS clouds: a Survey on Security in Multitenant Software Platforms. Computers & Security 31(1), 96-108.

Sakamoto, T., H. Yamada, H. Horie, and K. Kono. 2012. Energy-Price-Driven Request Dispatching for Cloud Data Centers. IEEE 5th International Conference on Cloud Computing (CLOUD). 2012, 974-976.

Salesforce. Salesforce1 Platform. 2014. http://www.salesforce.com/eu/platform/overview/ (retrieved 02/03/2014).

Samanthula, Bharath, Elmehdwi, Yousef, Howser, Gerry, Madria, Sanjay. 2015. A secure data sharing and query processing framework via federation of cloud computing. Information Systems (48), 196-212.

Satyanarayanan, Mahadev, P. Bahl, R. Caceres, and N. Davies. 2009. The Case for VM-Based Cloudlets in Mobile Computing. IEEE Pervasive Computing 8, n.º 4 (Oct 2009), 14-23.

SECCRIT project. https://www.seccrit.eu/ (retrieved 14/03/2014).

Seibold, M., A. Wolke, M. Albutiu, M. Bichler, A. Kemper, and T. Setzer. 2012. Efficient Deployment of Main-Memory DBMS in Virtualized Data Centers, IEEE 5th International Conference on Cloud Computing (CLOUD). 2012, 311-318.

Sen, Soumya, Carlee Joe-Wong, Sangtae Ha, and Mung Chiang. 2013. A Survey of Smart Data Pricing: Past Proposals, Current Plans, and Future Trends. ACM Computing Surveys (ACM) 46, n.º 2 (#nov# 2013), 15:1-15:37.

Services. Services World Congress 2013. 2013. http://www.servicescongress.org/2013/ (retrieved 02/03/2014).

Sharkh, M.A., M. Jammal, A. Shami, and A. Ouda. 2013. Resource allocation in a network-based cloud computing environment: design challenges. IEEE Communications Magazine 51, n.º 11 (November 2013), 46-52.

Sherry, Justine, Shaddi Hasan, Colin Scott, Arvind Krishnamurthy, Sylvia Ratnasamy, and Vyas Sekar. 2012. Making Middleboxes Someone else's Problem: Network Processing As a Cloud Service. Proceedings of the ACM SIGCOMM 2012 Conference on Applications, Technologies, Architectures, and Protocols for Computer Communication. New York, NY, USA: ACM, 2012, 13-24.

Shieh, Alan, Srikanth Kandula, Albert Greenberg, Changhoon Kim, and Bikas Saha. 2011. Sharing the Data Center Network. Proceedings of the 8th USENIX Conference on Networked Systems Design and Implementation. Berkeley, CA, USA: USENIX Association, 23-23.

Shin, S., Gu, G.. 2012. CloudWatcher: Network security monitoring using OpenFlow in dynamic cloud networks (or: How to provide security monitoring as a service in clouds?). In Proceedings of the 2012 20th IEEE International Conference on Network Protocols (ICNP), ser. ICNP '12, 1-6.

Silva, Joao, Moura, Jose, Marinheiro, Rui, Almeida, Joao. 2013. Optimizing 4G Networks with Flow Management using An Hybrid Broker. Proc. of Int. Conf. on Advances in Information Technology and Mobile Communication, Elsevier, 290-298.

SNIA. Cloud Storage Initiative. SNIA. 2014. http://www.snia.org/forums/csi (retrieved 03/03/2014).

Sridhar_a, T.. 2009. Cloud Computing - A Primer, Part 1: Models and Technologies. The Internet Protocol Journal, vol. 12, no .3., September of 2009. http://www.cisco.com/web/about/ac123/ac147/archived_issues/ipj_12-3/123_cloud1.html, 2-19.

Sridhar_b, T.. 2009. Cloud Computing - A Primer, Part 2: Infrastructure and Implementation Topics. The Internet Protocol Journal, vol. 12, no. 4., September of 2009. http://www.cisco.com/web/about/ac123/ac147/archived_issues/ipj_12-4/124_cloud2.html, 2-17.

Stallings, W.. 2015. Gigabit Ethernet: From 1 to 100 Gbps and Beyond. The Internet Protocol Journal, vol. 18, no. 1. March of 2015. http://ipj.dreamhosters.com/wp-content/uploads/2015/03/ipj18.1.pdf, 20-32.

Sterbenz, James, Hutchison, David, Çetinkaya, Egemen, Jabbar, Abdul, Rohrer, Justin, Schöller, Marcus, Smith, Paul. 2010. Resilience and survivability in communication networks: Strategies, principles, and survey of disciplines. Comput. Netw. 54 (8), 1245-1265.

Strijkers, Rudolf, Makkes, Marc, Laat, Cees de, Meijer, Robert. 2014. Internet factories: creating application-specific networks on demand. Computer Networks ( 68), 187-198.

Strunk, A.. 2010. QoS-Aware Service Composition: A Survey. Web Services (ECOWS), 2010 IEEE 8th European Conference on. 2010, 67-74.

Subashini, S., and V. Kavitha. 2011. Review: A Survey on Security Issues in Service Delivery Models of Cloud Computing. Journal of Network and Computer Applications (Academic Press Ltd.) 34, n.º 1 (#jan# 2011), 1-11.

Tanase, Mihai, and Valentin Cristea. 2011. Quality of Service in Large Scale Mobile Distributed Systems Based on Opportunistic Networks. Proceedings of the 2011 IEEE Workshops of International Conference on Advanced Information Networking and Applications. Washington, DC, USA: IEEE Computer Society, 2011, 849-854.

Tzanakaki, A., M.P. Anastasopoulos, G.S. Zervas, B.R. Rofoee, R. Nejabati, and D. Simeonidou. 2013. Virtualization of heterogeneous wireless-optical network and IT infrastructures in support of cloud and mobile cloud services. IEEE Communications Magazine 51, n.º 8 (August 2013), 155-161.

Vaquero, L.M., Rodero-Merino, L., Moran, D.. 2011. Locking the sky: a survey on IaaS cloud security. Computing 91(1), 93-118.

Vestin, J., Dely, P., Kassler, A., Bayer, N., Einsiedler, H., Peylo, C.. 2013. CloudMAC: towards software defined WLANs, SIGMOBILE Mob. Comput. Commun. Rev., 16(4), 42-45.

Vmware NSX. 2014. https://www.vmware.com/products/nsx/ (retrieved 17/03/2014).

Vogels, Werner. 2008. A Head in the Clouds: the Power of Infrastructure as a Service. Proc. of the 1st Workshop on Cloud Computing and Applications (CCA). 2008, 10 pages.

Voorsluys, William, James Broberg, and Rajkumar Buyya. 2011. Introduction to Cloud Computing. In Cloud Computing, 1-41. John Wiley & Sons, Inc., 2011.

Wang, Yating, Chen, Ing-Ray, & Wang, Ding-Chau. 2015. A Survey of mobile cloud computing applications: Perspectives and challenges. Wireless Personal Communications, 80(4), 1607–1623.

Ward, Jonathan, and Adam Barker. Undefined By Data: A Survey of Big Data Definitions. Arxiv.org. 20/09/2013. http://arxiv.org/pdf/1309.5821v1.pdf (retrieved 02/03/2014).

Wei, Yongjian, Guo, Junhu, Li, Hui, Ji, Yuefeng. 2014. Experimental demonstration of centralized control mechanism over all-optical network based on OpenFlow protocol, Optical Fiber Communications Conference and Exhibition (OFC), 1-3.

Wickboldt, Juliano, Esteves, Rafael, Carvalho, Marcio, Granville, Lisandro. 2014. Resource management in IaaS cloud platforms made flexible through programmability. Computer Networks (68), 54-70.

Wood, Timothy, K. K. Ramakrishnan, Prashant Shenoy, and Jacobus van der Merwe. 2011. CloudNet: Dynamic Pooling of Cloud Resources by Live WAN Migration of Virtual Machines. Proceedings of the 7th ACM SIGPLAN/SIGOPS International Conference on Virtual Execution Environments. New York, NY, USA: ACM, 2011, 121-132.

Wu, H., et al.. 2010. Network security for virtual machine in cloud computing. 5th International Conference on Computer Sciences and Convergence Information Technology (ICCIT), 18-21.

Yu, Haifeng, Phillip B. Gibbons, and Suman Nath. 2006. Availability of Multi-object Operations. Proceedings of the 3rd Conference on Networked Systems Design & Implementation - Volume 3. Berkeley, CA, USA: USENIX Association, 2006, 14 pages.

Zhang, Qi, Lu Cheng, and Raouf Boutaba. 2010. Cloud computing: state-of-the-art and research challenges. Journal of Internet Services and Applications (Springer-Verlag) 1, n.º 1 (2010), 7-18.